\documentclass[preprint]{aastex}
\usepackage{epsfig,psfig}
\usepackage{rotating,longtable}

\shorttitle{Sculptor dSph Galaxy}
\shortauthors{Coleman et al.}

\begin{document}

\title{The Absence of Extra-Tidal Structure in the Sculptor Dwarf Spheroidal Galaxy}

\author{M. G. Coleman}
\affil{Research School of Astronomy \& Astrophysics, Institute of 
Advanced Studies, The Australian National University, Cotter Road, Weston 
Creek, ACT 2611, Australia}
\email{coleman@mso.anu.edu.au}

\author{G. S. Da Costa}
\affil{Research School of Astronomy \& Astrophysics, Institute of 
Advanced Studies, The Australian National University, Cotter Road, Weston 
Creek, ACT 2611, Australia}
\email{gdc@mso.anu.edu.au}

\and

\author{Joss Bland-Hawthorn}
\affil{Anglo-Australian Observatory, PO Box 296, Epping, NSW 2121, Australia}
\email{jbh@aaoepp.aao.gov.au}

\slugcomment{Version of 3 June, 2005}
\begin{abstract}
The results of a wide-field survey of the Sculptor dwarf spheroidal galaxy are presented.  Our aims were to obtain an accurate map of the outer structure of Sculptor, and to determine the level of interaction between this system and the Galaxy.  Photometry was obtained in two colours down to the magnitude limits of $V=20$ and $I=19$, covering a $3.1 \times 3.1$ deg${}^2$ area centred on Sculptor.  The resulting colour-magnitude data were used as a mask to select candidate horizontal branch and red giant branch stars for this system.  Previous work has shown that the red horizontal branch (HB) stars are more concentrated than the blue HB stars.  We have determined the radial distributions of these two populations and show that the overall Sculptor density profile is well described by a two component model based on a combination of these radial distributions.  Additionally, spectra of the Ca {\sc ii} triplet region were obtained for over 700 candidate red giant stars over the 10 deg${}^2$ region using the 2dF instrument on the Anglo-Australian Telescope.  These spectra were used to remove foreground Galactic stars based on radial velocity and Ca {\sc ii} triplet strength.  The final list of Sculptor members contained 148 stars, seven of which are located beyond the nominal tidal radius.  Both the photometric and spectroscopic datasets indicate no significant extra-tidal structure.  These results support at most a mild level of interaction between this system and the Galaxy, and we have measured an upper mass limit for extra-tidal material to be $2.3 \pm 0.6 \%$ of the Sculptor luminous mass.  This lack of tidal interaction indicates that previous velocity dispersion measurements (and hence the amount of dark matter detected) in this system are not strongly influenced by the Galactic tidal field.
\end{abstract}


\section{Introduction}
The dwarf spheroidal (dSph) satellites of the Galaxy provide an important test of dark matter dynamics.  Simulations of heirarchical merging indicate that dwarf galaxies were the first structures to form in the early universe, and the merging and accretion of these objects created the galaxies and clusters of galaxies which exist today.  It is the collapse of dark matter which has driven the development of large-scale structure.  Observations indicate that the local dSph galaxies are potent repositories of dark matter (see \citealt{m98} and references therein) where the luminous material resides at the centre of a large dark halo.  However, these measurements of mass-to-light ratio ($M/L$) assume the system is in virial equilibrium.  \citet{kuhn89} proposed that the interaction of a satellite within the tidal field of its host galaxy may result in an artificially inflated estimate of the virial mass.  However, the simulations of \citet{piatek95} imply that the central mass-to-light ratio is not substantially affected by tidal forces.  In contrast, \citet{kroupa97} and \citet{klessen98} have created models of satellite systems {\em without} dark matter yet with apparently similar characteristics to those observed in the Galactic satellites.  Thus, tidal disruption by the Milky Way may be producing the appearance of dark matter in dSph galaxies.  The subsequent analysis of the Draco dSph by \citet{klessen03} did not detect the predicted line-of-sight depth in this system, however a quantitative measurement of the gravitational influence of the Galaxy on its satellites is yet to be achieved.

Consequently, there have been attempts to detect evidence of interactions between large galaxies and their satellites in the nearby universe.  Perhaps the best example of this process is the Sagittarius dwarf galaxy (\citealt{ibata94}; \citealt{majewski03} and references therein), a satellite currently being torn apart by the gravitational potential of the Galaxy.  The simulations of \citet{law04} have reproduced the general appearance of this disrupted system, and they find it is on a highly eccentric polar orbit ($e\approx 0.75$).  However, the structure of a satellite during the initial interaction with its host is still not understood.  Numerical simulations by \citet{helmi01}, \citet{mayer01} and \citet{johnston02} predict that during the early phase of interaction with the Galaxy, energy is injected into the outer regions of the satellite, resulting in a slightly inflated structure.  Also, depending on the level of interaction, the internal structure may be distorted.  An example of this initial stage may be the Ursa Minor dSph, which has a structure that displays significant asymmetries and extra-tidal stars \citep{mart-del01,palma03}.

The Sculptor dSph (Scl) lies at a Galactocentric distance of 80 kpc \citep{m98}.  Previous studies have attempted to detect the initial stage of Galaxy-satellite interaction by examining the radial profile of the Sculptor system.  For example, wide-field photometric surveys in a single colour \citep{eskridge88a,eskridge88b,ih95,walcher03} have identified possible substructure and found evidence for extra-tidal stars in this dSph.  Indeed, \citet{walcher03} claim the possible detection of tidal tails aligned with the Scl major axis.  However, the effect of foreground and background sources becomes important in the outer regions of this system due to the relatively small number of Scl stars compared to foreground Galactic stars.  As such, the existing radial profiles are highly uncertain in the outer regions, and it has therefore been difficult to accurately measure any possible inflation of the dSph structure.  Without a clear understanding of the level of Milky Way-Scl interaction, and hence whether this system is in virial equilibrium, it is not possible to accurately constrain the amount of dark matter in Scl using traditional velocity dispersion measurements.  Radial velocity studies by \citet{arman86} and \citet{queloz95} of K giants found a central mass-to-light ratio of $13 \pm 6$.  If the system is undergoing substantial distortion due to the Galactic tidal field, then this mass-to-light ratio may be overestimated.

We have searched the Sculptor dSph for structural and kinematic distortion.  Firstly, we collected wide-field photometry in two colours covering a 10 deg${}^2$ area centred on Scl, well beyond the nominal tidal radius of this galaxy.  The colour-magnitude diagram (CMD) obtained from the inner parts of the system was used to select candidate Scl stars over the dataset area, thereby increasing their contrast relative to field sources in the outer parts.  For examples of this process, see the analyses by \citet{grill95}, \citet{mart-del01}, \citet{piatek01} and \citet{majewski02}.  In the second stage, we have obtained spectra for $\sim$700 of the brightest candidate members.  This allowed a further refinement of the dataset by selecting stars based on radial velocity and metal abundances.  This paper describes our search for substructure in Sculptor based on the combined photometric/spectroscopic dataset.  A companion paper (\citealt{coleman05}; hereafter C05) presents the results for a similar wide-field photometric survey of Fornax.

\section{Photometric Survey} \label{photsurvey}

The observational strategy for this system followed that of the Fornax dSph (C05).  Images of the Sculptor region were obtained at the Siding Spring Observatory 1 metre telescope equipped with the Wide Field Imager (WFI).  Sixteen fields were observed in two filters ($V$ and $I$), thereby covering a $3.1^{\circ} \times 3.1^{\circ}$ area on the sky using a $4 \times 4$ mosaic of WFI fields.  Similar to the Fornax observations, accurate photometry was required to magnitudes of $V=20$ and $I=19$, encompassing the range of the Sculptor red giant branch (RGB).  A full list of observations, including the atmospheric seeing, is given in Table \ref{sculptorobservations}.  The data were reduced using the IRAF\footnote{IRAF is distributed by the National Optical Astronomy Observatories, which is operated by the Association of Universities for Research in Astronomy, Inc., under contract with the National Science Foundation.} program, and the stellar magnitudes were measured using the DAOPHOT program within IRAF.  An astrometric calibration of this dataset was made using the UCAC1 catalogue \citep{zach00}.  A full description of the data reduction and photometric calibration techniques can be found in C05.  

Multiply-detected stars in the $V$ and $I$ maps were removed by matching sources in the overlap regions with a search radius of $0.5''$.  The same search algorithm was used to combine the $V$ and $I$ datasets, thus producing a $(V-I)$ colour for all stars in common between the two maps.  Fig.\ \ref{sclcmd_cand} shows the resulting CMD from the region within $30'$ of the centre of Sculptor.  To measure the photometric completeness of the survey, a luminosity function was generated for each of the sixteen Sculptor fields in both $V$ and $I$ bands.  We then assumed the completeness limit of each field to be 0.2 mag brighter than the turnover in the luminosity function.  To ensure this method provided an accurate measurement of the completeness limit, we created a dataset of $\sim$5000 artificial stars with a distribution resembling the Scl RGB.  These stars were placed in each field, and then recovered using the photometry routines described above.  It was found that $94\%$ to $96\%$ of the artificial stars were recovered from each field to their respective photometry limits.  Thus, the luminosity function allowed an adequate measurement of the photometric limit for each field.  The overall magnitude limit of the survey is then defined by the field with the least depth: field Scl5 which is complete to $I=19$ and $V=20$.

\section{Spectroscopic Survey} \label{specsurvey}

\subsection{Observations}
The 2dF instrument at the prime focus of the Anglo-Australian Telescope (AAT) allows the simultaneous spectroscopy of up to 400 targets over a field of view $2^{\circ}$ in diameter.  A full description of 2dF is given by \citet{lewis02}.  Our aim was to collect spectra at 850 nm for candidate Scl stars, and to use the Ca {\sc ii} triplet spectral features to the separate Scl members (the `signal') from foreground/background sources (the `noise').  We chose 1408 candidate Scl stars from the photometric survey (described in the previous section) using the CMD selection range displayed in Fig.\ \ref{sclcmd_cand}.  Spectra were obtained for 893 of these candidates, however only 764 yielded spectra with the required signal-to-noise.  The majority of the stars for which spectra were not obtained lie close to, or within, three core radii, where the limit on fibre-to-fibre spacing of 2df restricts the object selection.  For the 527 stars in the sample beyond the tidal radius, however, $62\%$ have acceptable spectra.  The faint limit of this spectroscopic survey was set to be $I=17.5$ (approximately $V=18.6$ at the colour of the RGB, well above the photometric completeness limit given in the previous section) thus sampling the upper 2.0 mag of the Scl RGB.  The spatial distribution of these stars is shown in Fig.\ \ref{sclxy_cand}, where the five dashed circles outline the observed 2dF fields.  Note that we have observed potential member stars well beyond the Scl nominal tidal radius.  Approximately 100 stars located in the overlap regions between 2dF fields were observed more than once to ensure we obtained consistent velocities and Ca {\sc ii} line strengths.

The selected stars were observed with the 1200R gratings, and the resulting spectra were centred near the Ca {\sc ii} IR triplet with a wavelength range of $8100-9100$\AA \  and a scale of 1.07\AA \ per pixel.  Although the red spectral region is dominated by sky emission lines, we reduced such contamination using the slow-beam switching method\footnote{http://www.aao.gov.au/2df/skysub/skysub.html}, which intersperses object exposures with offset sky frames containing the same fibre configuration; that is, object and sky were observed through the same fibre.  In addition, each configuration included 42 fibres assigned to sky positions.  Hence, the observation sequence for each Scl field consisted of $6 \times 20$min exposures and three offset sky images.  Each sequence also included three wavelength calibration frames (arcs) comprising four Cu-Ar and two Cu-He lamps with an exposure time of 30s.  An arc frame at the start, middle and end of the sequence accounted for possible instrumental wavelength shifts.  Approximately $3-4$ hours were required to completely observe each field.  A list of observations is given in Table \ref{2dfobs}.

We also obtained spectra for 16 and 13 red giants in the globular clusters NGC 1904 and NGC 2298 respectively to be used as radial velocity standard stars.  Additionally, the metallicities of these clusters (NGC 1904, [Fe/H]$= -1.57$; NGC 2298, [Fe/H]$= -1.85$; from \citealt{harris96}) are approximately equal to that of Scl ($-2.8 \le$ [Fe/H] $\le -1.0$; \citealt{tolstoy04}).  For old, metal poor stars (such as those in Scl and the globular clusters) the individual equivalent widths of the calcium lines can be related to the metal abundance, and we can use the globular cluster observations to calibrate this relation.  Thus, additional to selecting stars based on radial velocity, the Ca {\sc ii} line strengths provided a second constraint to select members of the Scl system.

\subsection{Data Reduction and Analysis}
The spectra were reduced using the 2dfdr program\footnote{http://www.aao.gov.au/2df/software.html{\#}2dfdr} in an interactive mode.  In the 2dF setup, 400 spectra are divided between two CCDs, including 21 sky fibres per CCD.  As a first step in the data reduction process, a tram-line map was generated for both CCDs.  This map tracked the output spectra on each CCD, and was obtained from the fibre flat field images due to their high S/N.  The tram line map was then shifted and rotated to fit each individual data frame and the signal for each fibre extracted.  The arc frames then provide a wavelength calibration for each fibre by matching the pixel position of the observed lines with the spectral line list.

An accurate sky subtraction is dependent on a reliable fibre throughput calibration, since the relative throughputs of individual fibres can vary by $\sim$$10\%$.  The offset sky exposures were used to measure the relative fibre throughput.  These were taken with the same configuration as the object field, and multiple offset sky exposures at dithered positions ensured each fibre was accurately sampling the sky.  The offset sky frames were median combined to ensure the sky spectrum was not accidentally contaminated by sources or cosmic rays. The Scl frames were then reduced and fit-extracted, and the combined sky spectrum (from the 21 sky fibres), scaled by the individual fibre throughputs, was subtracted to remove the night-sky features.  The six exposures for each field were then median-combined to produce the final set of spectra.

\subsubsection{Radial Velocities} \label{radvels}
Due to less than favourable observing conditions (seeing $> 2''$), only 14 of the 29 globular cluster (GC) stars in the observed configurations had spectra with sufficient signal-to-noise ratio to be used as radial velocity standard stars.  Six of these were recorded on CCD 1 and eight on CCD 2.  To ensure these spectra were adequate templates, we used them to independently measure the mean systematic velocity of the two clusters, which were then compared to the known values.  The stellar radial velocities were measured using the {\em rvidlines} routine in IRAF.  Each velocity was calculated as the mean value determined from the three Ca {\sc ii} lines.  These were corrected for the motion of the Earth and we removed the CCD-dependence of these velocities using the $\Delta v_r$ values determined below.  Thus, we calculated a mean radial velocity of $205.3 \pm 2.2$ km s${}^{-1}$ for NGC 1904 (from six stars) and $152.7 \pm 1.4$ km s${}^{-1}$ for NGC 2298 (from seven stars, after removing a star whose radial velocity of 71 km s${}^{-1}$ indicated it is not a cluster member) where the errors are the standard deviation of the mean.  These compare well to the values listed in the online globular cluster catalogue (\citealt{harris96}; $206.0 \pm 0.4$ and $148.9 \pm 1.2$ km s${}^{-1}$ respectively).

In order to use these spectra as radial velocity standard spectra, we first needed to measure the velocity offset between each GC star.  For each CCD, the star with the highest S/N spectrum was defined to be the master star, and the remaining GC spectra were cross-correlated with the master spectrum to measure velocity offsets.  This resulted in six (eight) template spectra on CCD 1 (CCD 2) with a known velocity offset from the master star.  Radial velocities for the candidate Scl stars were then measured by cross-correlating the spectrum with all templates from the appropriate CCD, and shifting the resulting velocity by the appropriate offset such that they were measured relative to the master star for that CCD.  The spectra were cross-correlated in the wavelength range $\lambda \lambda 8380-8820$ \AA, which encompasses the Ca {\sc ii} triplet lines ($\lambda$8498, $\lambda$8542 and $\lambda$8662 \AA) and reduces the number of strong sky features which may not have been completely removed from the stellar spectra (see Fig.\ 1 in \citealt{arman86}) .  This was accomplished using the {\em fxcor} routine in the IRAF radial velocity package.  

Consequently, each star whose spectrum was recorded on CCD 1 (CCD 2) was cross-correlated with six (eight) GC spectra, resulting in multiple radial velocity measurements.  Since the velocities of the candidate Scl stars were measured relative to the two master stars, an adjustment was required to transform these to heliocentric velocities.  The heliocentric velocities of both master stars were previously measured using the {\em rvidlines} routine as described above.  Also, we calculated the heliocentric conversion for the Scl frames.  The difference between these values was added to the velocities of the candidate Sculptor stars, resulting in a set of radial velocity measurements relative to the Sun.  We combined these values to give a mean radial velocity for each potetial Scl member, and the velocity uncertainty was calculated as the standard deviation from the mean.

A final velocity zeropoint shift was achieved by comparing our measured radial velocities with those in a previous catalogue, which includes the observations of \citet{arman86}.  High resolution spectra for 38 stars were recorded over three seasons ($1985-1987$) at the CTIO 4 metre telescope, where the stars were selected from the \citet{hodge65} and \citet{kunkel77} photographic surveys of Scl.  \citet{arman86} described the data reduction and velocity measurement techniques and estimated the radial velocities had a mean error of $\sim$4 km s${}^{-1}$.  The current survey included 27 stars in common with this catalogue, where 14 spectra were recorded on CCD 1 and 13 on CCD 2.  From these, the mean velocity difference between this program and the \citet{arman86} catalogue was found to be,
\begin{displaymath}
\begin{array}{rl}
\mbox{CCD 1:} & \Delta v_r = 5.3 \pm 2.1 \mbox{ km s${}^{-1}$},\\
\mbox{CCD 2:} & \Delta v_r = 11.2 \pm 2.3 \mbox{ km s${}^{-1}$},\\
\end{array}
\end{displaymath}
\noindent where the errors are the standard deviation of the mean.  These $\Delta v_r$ values were added to the measured velocities to produce a final radial velocity for 764 stars.

To determine the mean uncertainty of the Scl radial velocities, we compared the measurements for multiply observed stars.  During the 2dF observing run, multiple observations were obtained of 120 stars located in the overlap regions between 2dF fields.  Of these stars, 92 yielded multiple radial velocities with an uncertainty less than 20 km s${}^{-1}$, and these cover the full magnitude range of the survey.  Approximately 60 of these were observed on both CCDs, and over half were obtained with significantly different fibre numbers.  This allowed cross-CCD comparisons between spectra recorded at different positions in each CCD.  The CCD velocity corrections listed above were applied to these values, and a comparison of these multiple observations yielded a mean velocity difference of $2.3$ km s${}^{-1}$, with a standard deviation of $14.9$ km s${}^{-1}$.  Thus, dividing this value by $sqrt{2}$, the Scl velocities have a mean uncertainty of approximately 10 km s${}^{-1}$.

Some of the measured velocities contained a large uncertainty, hence those with a velocity error greater than 20 km s${}^{-1}$ were removed.   This criterion removed 41 stars, resulting in a final catalogue containing 723 candidate Scl stars.  The majority of the rejected stars were located towards the faint end of the 2dF survey limit, where the mean velocity uncertainty was found to be $\sim$12 km s${}^{-1}$, as compared with $\sim$2 km s${}^{-1}$ at the bright end of the survey.  Fig.\ \ref{sclxy_cand} shows the spatial distribution of the final catalogue 723 stars.

\subsubsection{Line Strengths} \label{linestr}
The lines comprising the Ca {\sc ii} triplet are the strongest features in the far red component of a stellar spectrum, and are also situated close to the peak of the spectral energy distribution displayed by red giant stars.  Therefore, the pseudo-equivalent widths, or line strengths, of these three lines ($W_{\mbox{\scriptsize 8498}}$, $W_{\mbox{\scriptsize 8542}}$ and $W_{\mbox{\scriptsize 8662}}$) in RGB stars can be easily measured by medium-resolution spectroscopy.  A detailed description of the calcium triplet lines and their relation to red giant metallicities is given by \citet{cole04}.

\citet{arman91} first noted that if $\Sigma \mbox{Ca} = W_{\mbox{\scriptsize 8542}}+W_{\mbox{\scriptsize 8662}}$ is plotted against $V-V_{\mbox{\scriptsize HB}}$ (where $V_{\mbox{\scriptsize HB}}$ is the magnitude of the horizontal branch) for several stars in numerous globular clusters, then the resulting slopes are independent of cluster abundance.  This relation has been further investigated by \citet{suntzeff93}, \citet{dacosta95}, \citet{rutledge97a}, \citet{rutledge97b} and \citet{cole04}.  We refer the reader to Fig.\ 2 in \citet{dacosta95}, whose study of several Galactic globular clusters demonstrated that the best fitting line for the $(\Sigma \mbox{Ca},V-V_{\mbox{\scriptsize HB}})$ relation has a slope of $-0.62$ \AA/mag.  Consequently, this linear relationship can be used to remove the dependence of line strength on surface gravity by defining the reduced equivalent width to be $W' = \Sigma \mbox{Ca} + 0.62(V-V_{\mbox{\scriptsize HB}})$.  An important feature of $W'$ is that it can be directly related to the iron abundance \citep{arman91}.  \citet{dacosta95} analysed several Galactic globular clusters with a large abundance range and demonstrated that the $W'$-[Fe/H] relation can be split into two linear regimes.  This relation was further investigated by \citet{rutledge97b}, who found a similar result (albeit with a slightly different definition of the reduced equivalent width) using data for 52 Galactic globular clusters.  Therefore, the strengths of the Ca {\sc ii} lines provide an independent measure of metallicity for old, metal-poor stars.

In this survey, equivalent widths of the Ca {\sc ii} lines at $\lambda$8452 and $\lambda$8662 \AA \ were measured using the method described by \citet{arman91}.  Two windows to either side of each feature were used to trace the continuum at the feature wavelength.  The equivalent widths were then measured using a Gaussian fitting technique.  The same rest wavelength windows given in Table 2 of \citet{arman91} were used to fit the Gaussian functions and continuum line, however for the blue continuum bandpass of the Ca {\sc ii} $\lambda 8662$ line we used the wavelength range $\lambda \lambda 8559 - 8595$ \AA.  The weakest of the Ca II triplet lines, centred at a rest wavelength of $\lambda$8498 \AA, generally does not have sufficient S/N in our spectra, especially at low metallicities, for accurate measurement of the equivalent width.  Hence this line was not included in the following line strength analysis.

Fig.\ \ref{ewgc}(a) shows the sum of the two strongest Ca {\sc ii} line strengths as a function of magnitude difference from the horizontal branch for the two globular clusters NGC 1904 (closed circles) and NGC 2298 (open circles).  We have discounted one of the stars in the NGC 2298 dataset, since its velocity and equivalent width data indicate it is not a member of the cluster.  Also, two of the stars have been removed from the set of NGC 1904 points, as their $\Sigma \mbox{Ca}$ values contained large uncertainties and differed significantly from the remaining four stars.  The dashed lines in Fig.\ \ref{ewgc}(a) trace the best linear fit to the data with a slope of $-0.62$ \AA/mag.  This slope provided an adequate linear fit to the data points for both clusters.

To ensure these results are consistent with previous observations, we calculated the mean reduced equivalent width ($\overline{W'}$) for both clusters.  \citet{rutledge97b} list values for the two clusters which can be easily modified to the current system.  From the transformation derived by \citeauthor{rutledge97b}, we find their $\overline{W'}$ values for the clusters NGC 1904 and 2298 to be $3.24 \pm 0.12$ and $2.32 \pm 0.05$ \AA \ respectively.  We calculated $W'$ for each star using the data shown in the top panel of Fig.\ \ref{ewgc}, and found the following mean reduced equivalent width for each cluster,
\begin{displaymath}
\overline{W'} = \left \{
 \begin{array}{ll}
3.23 \pm 0.10 \mbox{\ \AA} & \mbox{(NGC 1904)} \\
2.51 \pm 0.17 \mbox{\ \AA} & \mbox{(NGC 2298)} \\
\end{array} \right.
\end{displaymath}
\noindent where the uncertainties are the standard deviation of the mean.  These are consistent with the \citet{rutledge97b} values.

A second verification of these reduced equivalent widths was to ensure they satisfy the empirical $\overline{W'}$-[Fe/H] relation derived by \citet{dacosta95} using the globular cluster metallicities provided by \citet{zinn84}.  The dashed line in Fig.\ \ref{ewgc}(b) illustrates this relationship for metal-poor clusters, and the two data points represent the $\overline{W'}$ values listed above.  The [Fe/H] values are from \citet{zinn84}, to ensure consistency with the \citeauthor{dacosta95}\ study.  The datapoints are adequately described by the abundance relation, hence the mean reduced equivalent widths derived for these two globular clusters are consistent with previous observations.

The overall uncertainty of the line strength measurements was obtained by comparing 71 multiply observed stars in the overlap regions between 2dF fields that have the same approximate luminosity distribution as the total set of Scl candidates.  For the $W_{\mbox{\scriptsize 8542}}$ and $W_{\mbox{\scriptsize 8662}}$ lines, we measured the standard deviation from the mean to be 0.31 \AA \ and 0.38 \AA \ respectively.  The total uncertainty for $\Sigma \mbox{Ca}$ was measured to be $\sigma_W = 0.42$ \AA.  This uncertainty is dominated by instrumental and sky noise, and is only weakly dependent on magnitude.

\section{Results -- Photometric Survey} \label{photresults}
The early photographic studies of \citet{hodge61,hodge65,hodge66} established many of the fundamental structural parameters of Scl, and did not detect any significant deviations from a smooth stellar density profile.  However, later analyses \citep{demers80,eskridge88a,eskridge88b} found that the region within the core radius is almost circular, yet the ellipticity increases with radius.  These analyses also detected possible extra-tidal stars.  More recently, \citet{ih95} conducted a structural analysis of Scl based on a `blue' (IIIaJ) photographic plate, and found that its structure closely resembles the numerical simulations of tidally distorted galaxies by \citet{piatek95}.  Irwin \& Hatzidimitriou also found some evidence for extra-tidal stars.  The CCD survey of \citet{walcher03} detected possible tidal tails extending along the major axis of Scl in the East-West direction and concluded that the outer structure of Scl is strongly influenced by the Galactic tidal field.  Consequently, to further address these claims, we have investigated the structure of Scl using a variety of methods applied to the photometric dataset.

The CMD provides a filter to select candidate Scl stars, thereby decreasing the number of field stars.  In a similar photometric survey of the Fornax dwarf spheroidal galaxy, we described a simple colour-magnitude selection technique which decreased contamination due to field stars (C05).  For the Scl survey we applied a more complex CMD-selection process based on that of \citet{grill95}, which has since been adapted by \citet{oden01}.  This method required the CMD to be divided into an array of cells, and provided a `signal' value of Scl stars throughout colour-magnitude space.  Those cells with a signal greater than the limiting value were then used to select candidate Scl stars, maximising the ratio of Scl to field stars in the outer regions of the dwarf galaxy.  A full description can be found in \citet{grill95}, however we include a summary of the process below, detailing how it was applied to our data.

To determine the signal of Scl stars over the CMD, we divided the photometric data into two spatial regions; the `core' region (within three core radii of the Scl centre), which is dominated by Scl stars; and the field star region (located outside the tidal radius and within the survey limit).  A grid was overlaid on the CMD, subdividing it into a series of cells with indices $(i,j)$.  Each cell had a width of 0.05 mag in colour and a height of 0.10 mag.  To minimise the effects of photometry errors, we only considered the magnitude range $16 \le V \le 20$, where the mean colour uncertainty is less than 0.03 mag.  For each colour-magnitude cell $(i,j)$, we counted the number of stars located in the core region $n_C(i,j)$, and the number in the field region $n_F(i,j)$.  Thus, the signal-to-noise ratio of core to field stars for each CMD cell is,
\begin{displaymath}
s(i,j) = \frac{n_C(i,j) - gn_F(i,j)}{\sqrt{n_C(i,j) + g^2n_F(i,j)}},
\end{displaymath}
\noindent where $g$ is the ratio of the core area to the field star area.  In effect, the array $s(i,j)$ listed which regions of the CMD contained a high signal-to-noise of Sculptor stars.  Following the method of \citet{grill95}, we defined a limiting signal value $s_{\mbox{\scriptsize lim}}$ to optimise the cumulative signal-to-noise ratio of core stars to field stars.  Stars were then selected in those CMD cells with a signal greater than a limiting signal value: $s(i,j) > s_{\mbox{\scriptsize lim}}$.  A contour plot showing the signal $s(i,j)$ over colour-magnitude space is given in Fig.\ \ref{sclcmdlines}.  The outer contour outlines those cells with a signal greater than $s_{\mbox{\scriptsize lim}}$.  This region encompasses the Scl RGB down to the photometric limit, and hence was used to select candidate RGB stars.

Thus, we selected 4355 candidate Scl RGB stars, and the spatial distribution is displayed in Fig.\ \ref{sclrgbxy}, where the ellipses represent the core and tidal radii of the RGB population (see Table \ref{kingpar}).  At first glance, there is no obvious substructure in the distribution of Scl RGB stars.  The radial profile of this stellar population is displayed in Fig.\ \ref{sclradial}.  The best-fitting \citet{king66} model to all datapoints is represented as a dashed line, from which we derived a nominal tidal radius of $72.5 \pm 4.0$ arcmin, which is comparable to the \citet{ih95} value of $r_t = 76.5 \pm 5$ arcmin.  However, there is some deviation from this curve beyond $r \approx 30'$.  Consequently, we fitted a King profile to the inner $40'$, which is represented by a dotted line.  This profile had a limiting radius of $40.5 \pm 2.5$ arcmin, and the data shown in Fig.\ \ref{sclradial} implies significant `extra-tidal' structure.  This is consistent with the results of \citet{walcher03} who found a tidal radius of $44'$ and claimed a large population of stars beyond this radius.  We have found that an over-subtraction of the field population in our data can reproduce the results of \citeauthor{walcher03}, giving a detection of extra-tidal structure.  It is clear that neither model provides an adequate description of the Sculptor system, however the King parameters for both models are listed in Table \ref{kingpar}, where the entries for `RGB-a' refer to the profile for the inner $40'$ of Scl.

In this context, it is worth considering the results of \citet{tolstoy04}, who found that the red giants of Sculptor apparently consist of two populations which are kinematically and chemically distinct.  These two populations are probably related to the fact that the red and blue HB stars in Sculptor have different radial distributions with the red HB stars being more centrally concentrated \citep{kaluzny95,dacosta96,harbeck01,majewski99,hurley99}.  Section 4.3 contains a full analysis of the blue and red HB star distributions, from which we have derived their spatial extents; the King parameters for both populations are given in Table \ref{kingpar}.  We have used these distributions to examine the overall density profile of the RGB stars, finding that a two-component fit gives an excellent representation of the data.  This is shown as the dashed line in Fig.\ \ref{tworadial}, where the dotted lines represent the two components.  These two components have the core and limiting radii listed for the blue and red HB populations respectively.  Also, to reduce the number of free parameters, the ratio of the central densities of these two components is the same as the blue HB/red HB ratio, taken from Table \ref{kingpar}.  The resulting model provides a significantly better fit to the data than a single component model.

\subsection{Extra-Tidal Structure} \label{outerstr}
\subsubsection{Density Probability Function}
In the companion paper (C05) we described two methods to detect structure beyond the nominal tidal radius of a stellar system.  The first of these, a density probability function, was created by measuring the stellar density in a large number of randomly placed circles over the survey area.  In this way, we were able to find the probability of measuring a given density of stars {\em outside} the nominal tidal radius.  A uniform distribution of stars (for example, the field star population) is expected to display a Gaussian density function, while substructure will produce a non-Gaussian function.  This function was measured for the distribution of RGB-selected stars outside the nominal tidal radius of Scl, where the stellar density was measured within $10^6$ circles of radius $12'$.  The circle size was chosen to be large enough to include a reasonable number of stars, yet not too large to miss the detection of medium-scale ($\sim$20 arcmin) structures.  Fig.\ \ref{mc_xy} shows the density function, where the dashed line represents the best-fitting Gaussian to the data.  This function indicates the mean background stellar density is 237 stars/deg${}^2$ with a standard deviation of 65.  Unlike the corresponding result for Fornax (C05), the extra-tidal region of Scl does not display any significant substructure; the function appears to be consistent with a uniform distribution of sources, which implies a population comprising only field stars.

If the stars beyond the tidal radius are not part of the Scl system, then they must be foreground and background sources.  The background source population is made up of distant galaxies, however the majority of these would not be interpreted as stellar sources by the DAOPHOT program due to their non-circular morphology.  Although a complete star-galaxy separation cannot be made given an average seeing of $2''$ for the survey, we estimate that the total number of background galaxies to be less than $25\%$ of the number of objects detected.  Also, the CMD selection range shown in Fig.\ \ref{sclcmdlines} further precludes this population.  Hence, assuming the extra-tidal population of Scl is negligible, the field population is almost entirely assembled from foreground Galactic stars.  \citet{rat85} estimated the density of the field star population towards Sculptor using the Galactic models of \citet{bahcall80}.  In the colour-magnitude range $17<V<21$, $0.8<(B-V)<1.3$, they estimated 252 stars/deg${}^2$ in the Scl field, with an approximate error of $25\%$.  Although these colour and magnitude limits do not exactly correspond with our CMD selection range (Fig.\ \ref{sclcmdlines}), the \citet{rat85} estimate agrees (within the uncertainties) with our mean field star density measurement of 237 stars/deg${}^2$ represented in Fig.\ \ref{mc_xy}.  This supports our inference that the candidate-RGB stars beyond the nominal tidal radius correspond to the field star population, and there is no substantive evidence for Scl RGB stars in this region.

To determine how effectively the density probability function can detect structures in an otherwise random distribution of stars, we simulated a series of overdensities in the extra-tidal region of Scl.  The Monte Carlo algorithm was applied to a dataset consisting of the candidate Scl RGB stars and two synthetic tidal tails extending in the North and South directions from the centre of the system.  We created six artifical datasets, where the mean stellar density of the synthetic tidal tails was $\rho = 0.5\sigma, 1.0\sigma, \dots, 3.0\sigma$ above that of the field population.  The tails uniformly occupied the region $-0.5^{\circ} \le \Delta \alpha \le 0.5^{\circ}$ over the total declination range of the survey, both inside and outside the nominal tidal radius.  For example, a spatial plot of the RGB dataset with artificial tidal tails of density $2.0\sigma$ can be seen in Fig.\ \ref{sclrgbtails}, which (from the aggregate stellar luminosity over this sky area) have a surface brightness of 30.4 mag/arcsec${}^2$ above that of the field population.

Fig.\ \ref{mc_xy_tot} shows the density functions measured from the six tidal tail simulations.  For each panel, the dashed line traces the density function measured from the true extra-tidal population.  The $0.5\sigma$ overdensity population is barely visible in the density function, however a second peak becomes more prominent as the density of the synthetic tidal tails increases.  For tidal tails of this type, we find evidence for substructure when the density is a factor of $\sim$$1.5\sigma$ above the background population.  This corresponds to $\sim$100 extra-tidal RGB stars per square degree, or approximately $10\%$ of the light from all Sculptor RGB stars.    The simulations were repeated using tidal tails of various widths (greater than $\sim$$0.3^{\circ}$) aligned at several position angles relative to Sculptor, and we found that the density probability function was able to detect features with a surface brightness greater than $\mu_V \sim 31$ mag/arcsec${}^2$ above the background population, assuming they have the same luminosity function as Scl.


\subsubsection{Angular Correlation Function}
A second method to search for extra-tidal structure is to measure the angular correlation function for the stellar population beyond the nominal tidal radius.  The angular correlation function is generally used in cosmology to examine if a sample of galaxies are clustered, or alternatively, follow a Poissonian distribution \citep{peebles80}.  For a given sample of data points covering a finite area, this function is measured by placing a large number of random points in a two-dimensional area with the same shape as the dataset.  The number of data-data pairs ($DD(\theta)$), data-random pairs ($DR(\theta)$) and random-random pairs ($RR(\theta)$) is then determined at a given scale length $\theta$.   The observed angular correlation function is then calculated as,
\begin{equation} \label{angcorobs}
w_{\mbox{\scriptsize obs}}(\theta) = 1 + \frac{DD(\theta)}{RR(\theta)}W_1 - 2 \frac{DR(\theta)}{RR(\theta)}W_2.
\end{equation}
\noindent The parameters $W_1$ and $W_2$ are determined from the number of data and random points ($N_{\mbox{\scriptsize data}}$, $N_{\mbox{\scriptsize ran}}$) using the relations,
\begin{eqnarray}
W_1 & = & \frac{N_{\mbox{\scriptsize ran}}(N_{\mbox{\scriptsize ran}}-1)}{N_{\mbox{\scriptsize data}}(N_{\mbox{\scriptsize data}}-1)}, \\
W_2 & = & \frac{N_{\mbox{\scriptsize ran}}-1}{N_{\mbox{\scriptsize data}}}.
\end{eqnarray}
\noindent This provides a solution for a survey of infinite extent, however for a finite survey the function must be adjusted by a constant called the integral constraint, defined to be the product $A_w B$ \citep{roche99}.  The parameter $B$ is determined numerically from the number of random-random pairs, and the values of $A_w$ and $\beta$ (defined below) are determined as those which provide the best-fitting function to the observed datapoints.  After correcting for the integral constraint, the angular correlation function takes the form,
\begin{equation} \label{angcorfn}
w(\theta) = A_w \theta ^{-\beta},
\end{equation}
\noindent where $A_w$ is the function amplitude, and $\beta$ defines the slope.  A purely random distribution of points will produce a flat angular correlation function ($A_w$ and $\beta$ are both equal to zero, hence $w(\theta)=0$), while a sloped function ($A_w$ and $\beta$ are greater than zero) reflects a clustered sample.

Accordingly, we measured the angular correlation function for all stars beyond the nominal tidal radius of Scl.  We placed $10^5$ points randomly throughout the survey region and measured the distance between all data-data, data-random and random-random pairs.  These were placed in logarithmic bins of width $\Delta \log \theta = 0.05$, where $\theta$ is in arcmin.  Fig.\ \ref{corrfn} shows the data points for the observed angular correlation function, calculated using Eqn.\ \ref{angcorobs}.  The errorbars were calculated from the Poisson noise of the number of data points in each bin of $\theta$.  The dashed line defines $w_{\mbox{\scriptsize obs}}(\theta) = 0$, which is the expected function for a randomly distributed set of stars.  This line accurately traces the data points, especially at large scales where a large number of data-data pairs in each bin results in small uncertainties.  Hence, this test indicates that there is no significant structure in the distribution of RGB-candidate stars beyond the nominal tidal radius of Scl.

To test the effectiveness of this function, we repeated the process for the six synthetic tidal tail datasets described in the previous section.  Fig.\ \ref{acftest} displays the resulting angular correlation functions, where the parameter at the top right corner of each panel is the density of the tidal tails in terms of the standard deviation of the field population ($\sigma$).  The errorbars represent Poisson noise and the dashed line traces $w_{\mbox{\scriptsize obs}}(\theta) = 0$, the expected function for a normal distribution of points.  Similar to the density function above, we have found an overdensity of $\sim$$1.5\sigma$ is easily discernible, while structures with a density less than this value are difficult to distinguish from the background noise.  Hence, the angular correlation function was able to detect extra-tidal structures with a surface brightness greater than $\mu_V \sim$31 mag/arcsec${}^2$.  That is, in order for an extra-tidal structure to have been detected around Scl, it required a surface brightness approximately $30\%$ greater than the background.

\subsubsection{Contour Map}
Both tests indicated there was no substructure in the extra-tidal distribution of stars shown in Fig.\ \ref{sclrgbxy}.  To further investigate the possibility of extra-tidal structure, we created a contour map of Scl (Fig.\ \ref{sclcontour}) from the RGB-selected stars.  Each star was convolved with a Gaussian of radius $40''$, and the contour smoothing length is $3'$.  The outer contour traces all structure with a density $1\sigma$ above the field population and the second contour level represents a $3\sigma$ density, where $\sigma = 65$ stars/deg${}^2$ is the standard deviation of the field population.  The first contour is highly sensitive to statistical variations in the field star density, hence we define the $3\sigma$ level as the significance limit of this survey.  The red ellipse in Fig.\ \ref{sclcontour} represents the Scl tidal radius.

There are some signs of structure in the outer regions of Sculptor.  For instance, the two small $3\sigma$ density peaks located in the south-west quadrant.  \citet{schweitzer95} measured the absolute proper motion of Scl to be directed towards the NE, thus it is possible that the south-west `blobs' represent tidally stripped stars trailing the Sculptor system.  It is also possible, however, that these structures are simply density spikes in the field star population, and are therefore not associated with Sculptor.  Either way, they represent only a small fraction of the stars in this system.

A second feature is the small `hook' structure centred at approximately $\Delta \alpha = -0.2$, $\Delta \alpha = -0.6$, within the tidal radius.  There is some evidence that this structure also exists in the \citet{walcher03} study, however only at the $1 - 2 \sigma$ level in their diagram.  It points towards the south-west quadrant, and possibly to the small $3\sigma$ density peaks mentioned above.  This structure may represent tidal distortion, or it may also be a random fluctuation in the field star density.

\subsection{Inner Structure} \label{innerstr}
We investigated whether the structure of Scl RGB stars is dependent on radius by fitting a series of ellipses to the stellar density, following the method described in C05.  The best-fitting centre, position angle and ellipticity were calculated at each major axis radius value of $r=5, 6, \dots , 60$ arcmin.  Beyond $r=60'$, the number of stars was too low to accurately fit an ellipse to the data.  Fig.\ \ref{lscontour} shows the mean values in radial bins of $5'$, where the uncertainties represent the standard deviation from the mean and the dashed lines mark the central coordinates, ellipticity and position angle listed by \citet{m98}.  The upper two plots demonstrate that the centre of Scl remains constant out to a radius of approximately $40'$, and then moves North for all subsequent radii.  Although the position angle of these ellipses effectively remains constant at all radii, the ellipticity plot in Fig.\ \ref{lscontour}(d) indicates that the structure of the inner regions is almost circular, and then becomes more elliptical as the radius increases (see Fig.\ \ref{sclcontour}).  This reinforces the results of \citet{demers80}, \citet{eskridge88a} and \citet{ih95}.

Hence, there is some evidence for Scl distortion.  A visual examination of the contour diagram in Fig.\ \ref{sclcontour} does indicate that the outer region has a significantly higher ellipticity compared to the inner region.  \citet{walcher03} reported the possibility of tidal extensions along the East-West axis (see their Fig. 2).  Although \citeauthor{walcher03}\ state a limiting magnitude for their survey to be $V \sim 23.5$, they also point out that the claim of tidal extensions is subject to non-uniform limiting magnitudes throughout the survey, and is also dependent on the level of star/galaxy separation.  We confirm that the structure of Scl appears stretched along this axis in the outer regions, however there is little evidence to support the existence of well-defined tidal tails.

\subsection{Horizontal Branch Stars} \label{hbstars}
\citet{kuhn89} proposed that the interaction between a satellite system and the host galaxy may artificially increase the velocity dispersion, thereby increasing the measured mass-to-light ratio.  However, in order to increase the radial velocity dispersion the satellite galaxy must be positioned such that its long axis is aligned along the line of sight.  \citet{klessen98} stated that the resulting distance distribution of the stars should produce a horizontal branch width of $\sim$1 mag.  \citet{klessen03} previously searched for this line-of-sight depth in the Draco dSph, without success.  In Scl we find the horizontal branch vertical dispersion to be approximately 0.15 mag which does not support a line-of-sight inflation of this system.  For the remainder of this section, we investigate the morphology of the Scl horizontal branch in relation to the second parameter effect.

The horizontal branch (HB) constitutes core He-burning stars with ages greater than $\sim$10 Gyr.  Studies of globular clusters indicate that metallicity is the first parameter governing the HB morphology; metal-poor populations tend towards a blue horizontal branch (BHB), while metal-rich populations often form the red horizontal branch (RHB).  However, some systems exhibit a second parameter (2P) effect, where the HB morphology is not wholly dependent on metallicity.  The early photometric analyses of Scl revealed a surprisingly red HB given its mean abundance \citep{kunkel77,dacosta84,dacosta88}, implying Scl is a 2P object.

More recently, deeper photometry over a wide field has revealed that the HB morphology in Scl is radially dependent \citep{light88,kaluzny95,dacosta96}, such that the RHB stars are more centrally concentrated than the BHB stars.  This has been verified by \citet{harbeck01}.  The results of \citet{majewski99} and \citet{hurley99} indicate that the stellar population of Scl is bimodal; that is, the inner region displays the 2P effect and contains a mixture of RHB and BHB stars, while the outer region is dominated by BHB stars.  The spectroscopic study by \citet{tolstoy04} showed that Sculptor contains two distinct stellar populations represented by either side of the HB, and that these populations are offset from each other in metallicity and kinematics.  The population gradient in Scl may indicate that residual gas from the first star formation burst was concentrated at the centre of the system, resulting in younger and/or more metal rich central stellar population (for a full discussion, see \citealt{harbeck01}).

\begin{figure}
\plotone{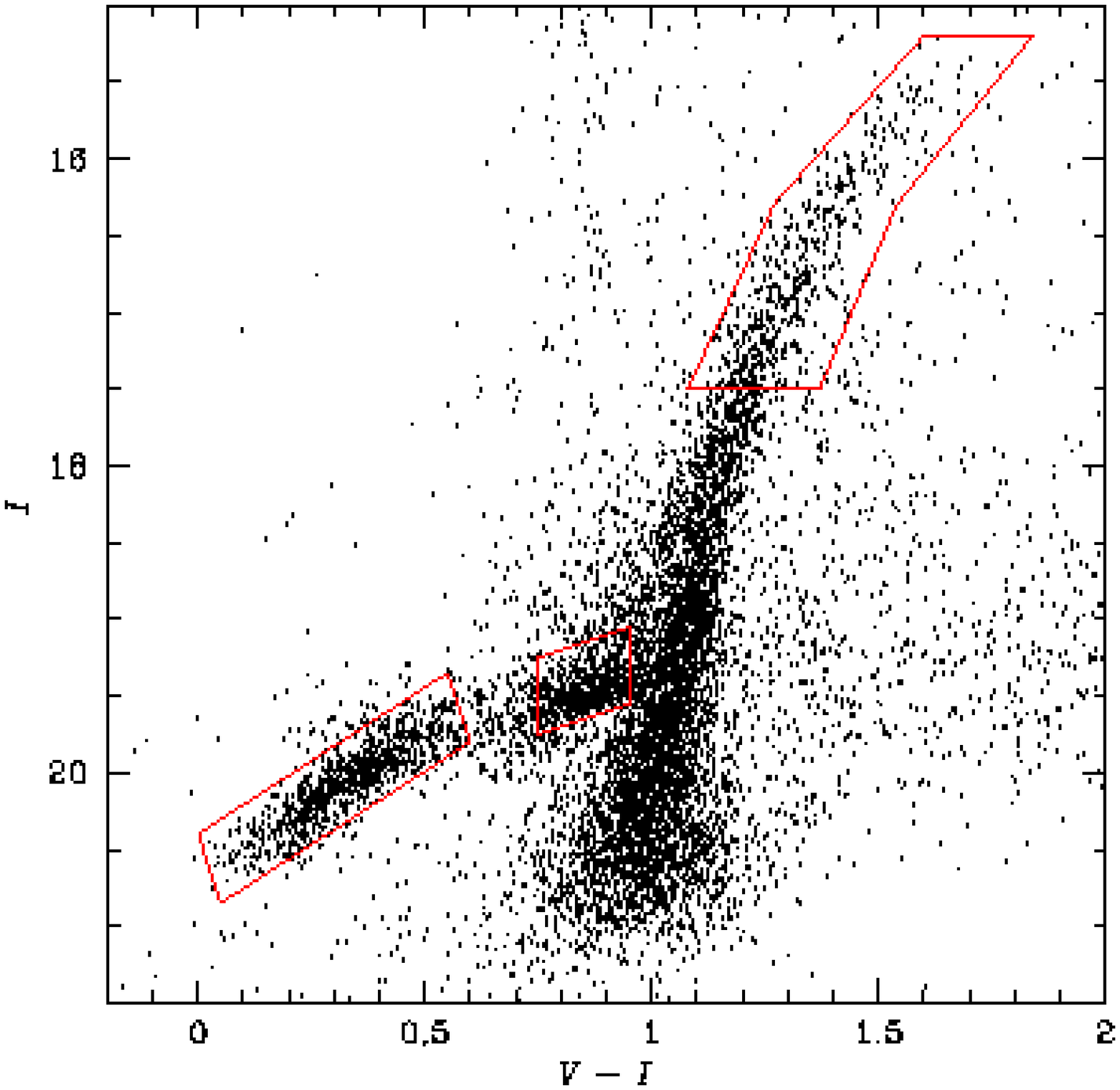}
\figcaption[CMD of the 2dF Scl candidate selection]{The Scl CMD.  The upper solid line encloses the colour-magnitude range of the Scl candidate stars observed with 2dF.  The two lower polygons outline the selection regions for the red and blue horizontal branch stars.  \label{sclcmd_cand}}
\end{figure}

\begin{figure}
\plotone{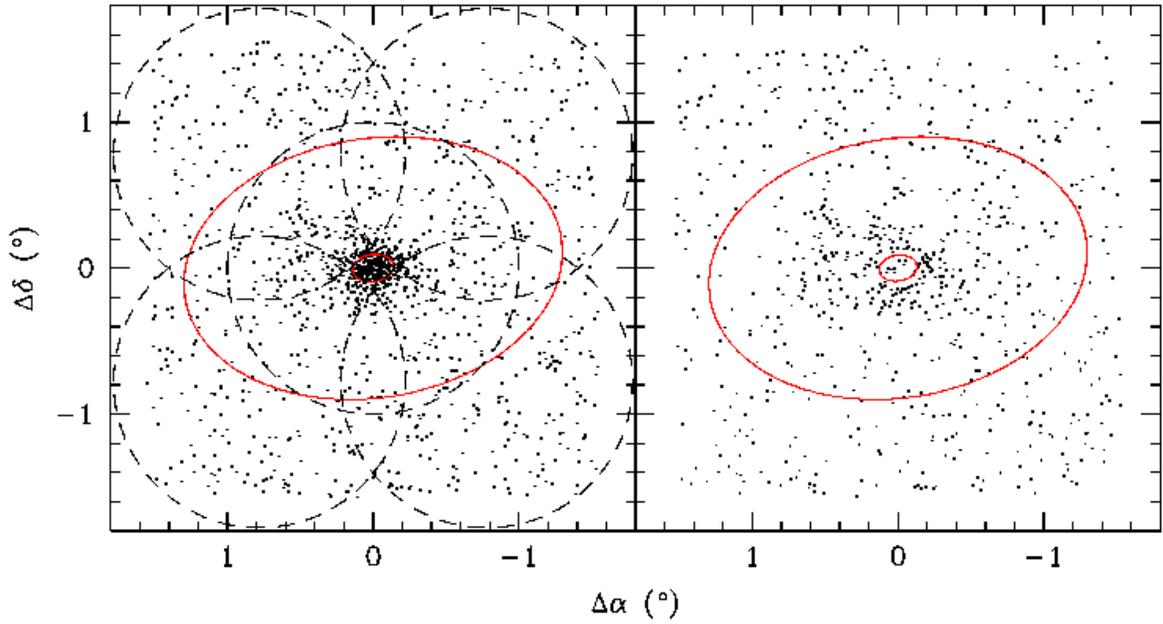}
\figcaption[Spatial distribution of 2dF Scl candidates]{{\em Left Panel:} Spatial distribution of the 1408 stars selected within the colour-magnitude range in Fig.\ \ref{sclcmd_cand}.  The inner and outer ellipses trace the core and nominal tidal radii measured from the RGB stars (see \S \ref{photresults}).  The five dashed circles are two degrees in diameter, and define the field placement for the 2dF observations. {\em Right Panel:} The 723 stars for which we obtained radial velocity errors less than 20 km s${}^{-1}$.  \label{sclxy_cand}}
\end{figure}

\begin{figure}
\plotone{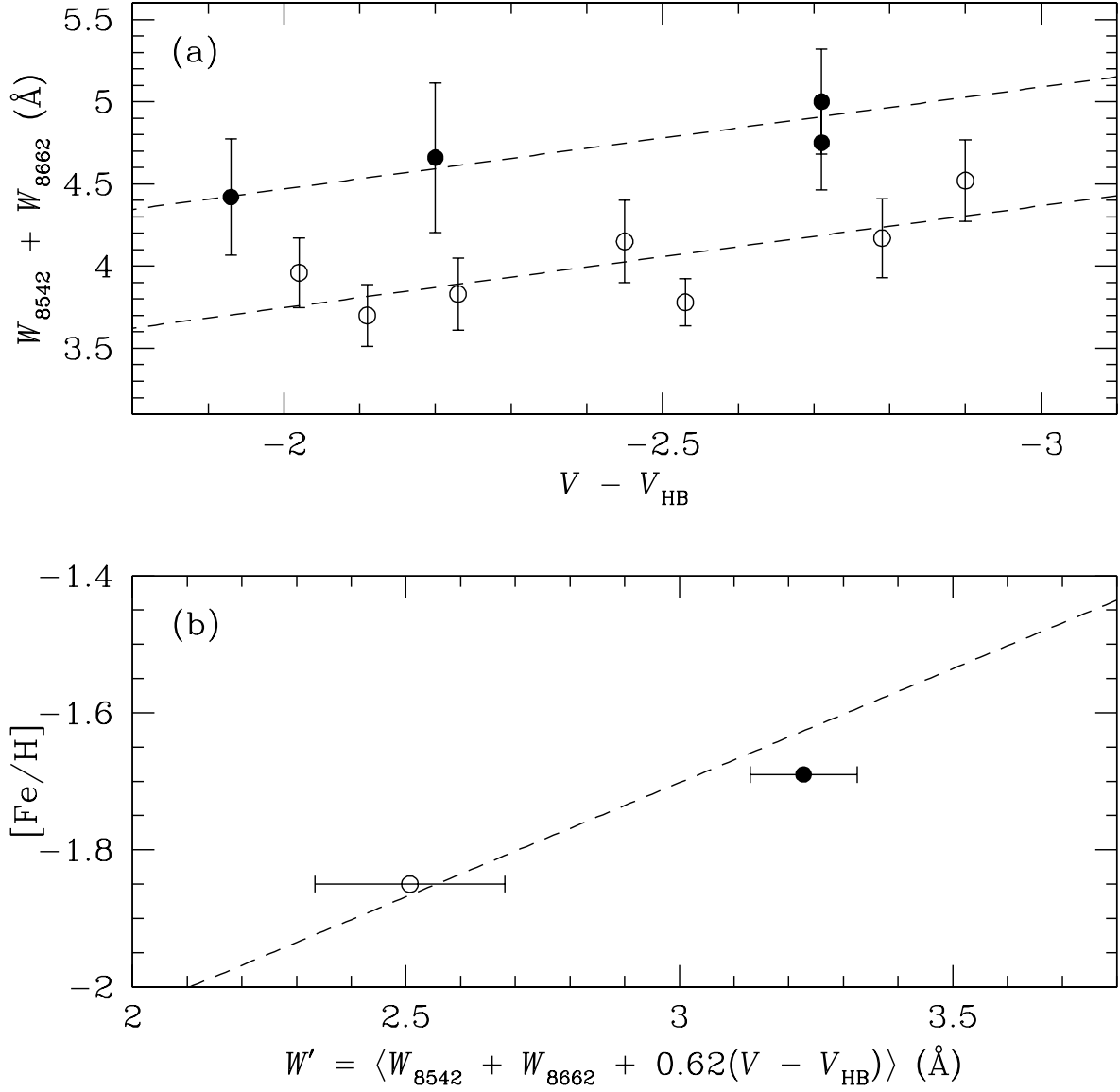}
\figcaption[Ca {\sc ii} line strength against $V-V_{\mbox{\scriptsize HB}}$]{(a) The Ca {\sc ii} line strength index $\Sigma \mbox{Ca}$ as a function of magnitude difference from the horizontal branch, $V-V_{\mbox{\scriptsize HB}}$.  The lines have a slope of $-0.62$ \AA/mag.  The open circles represent stars from NGC 2298, and the closed circles those NGC 1904.  (b) The mean reduced equivalent width, $\overline{W'}$ against metallicity for the two globular clusters using the same notation as in (a), where errorbars are the standard deviation from the mean.  The dashed line represents the metal-poor relation from \citet{dacosta95} used to calibrate the abundance of the Ca {\sc ii} line strengths.  \label{ewgc}}
\end{figure}

\begin{figure}
\plotone{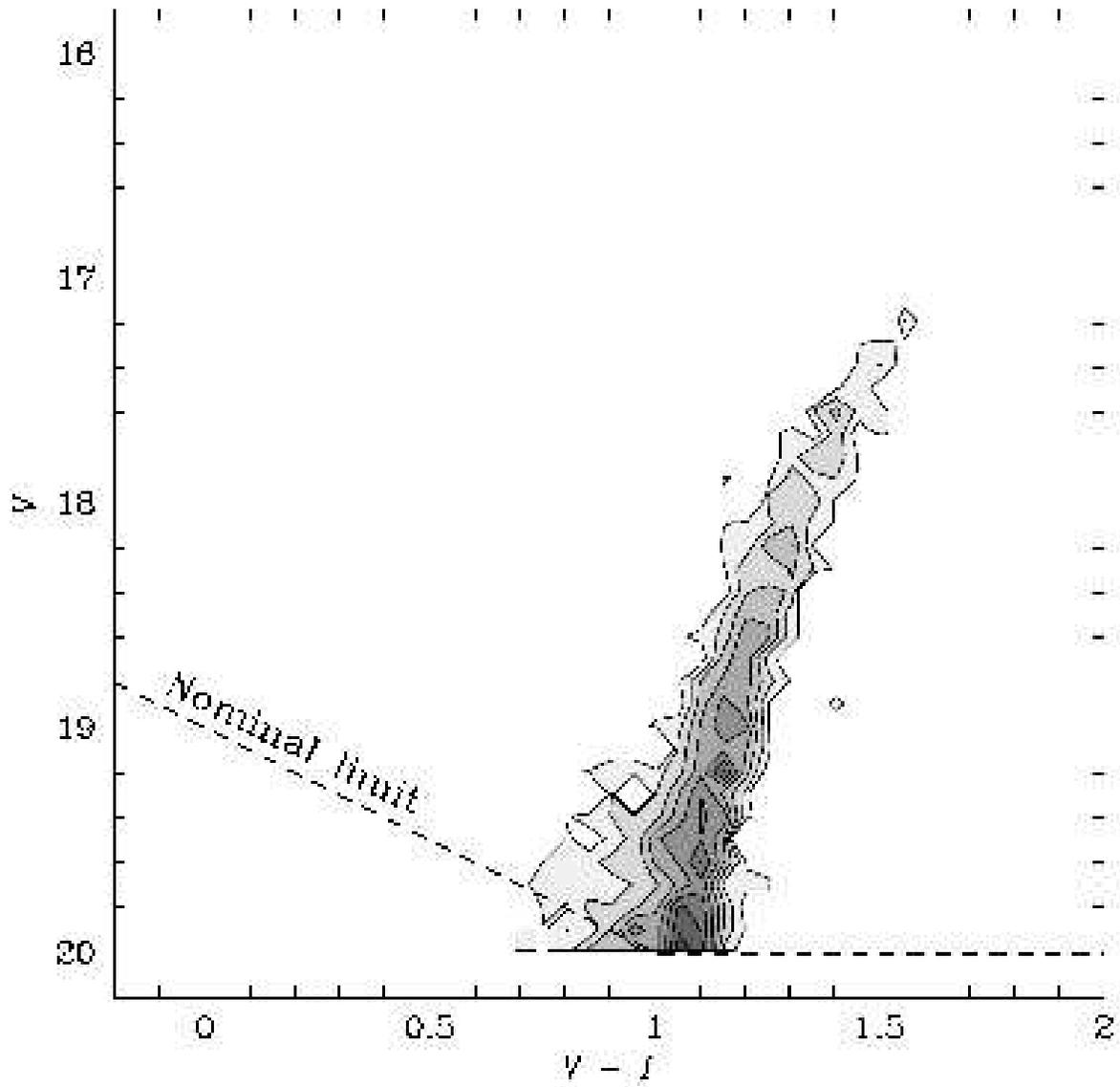}
\figcaption[Sculptor CMD selection]{CMD contour diagram of the signal-to-noise function $s(i,j)$.  The outer contour traces all colour-magnitude cells with a signal greater than $s_{\mbox{\scriptsize lim}}$, and thus outlines the CMD selection range for candidate Scl stars.  The region effectively encompasses the Scl red giant branch, where the dashed line represents the photometric limit of the entire dataset.  \label{sclcmdlines}}
\end{figure}

\begin{figure}
\plotone{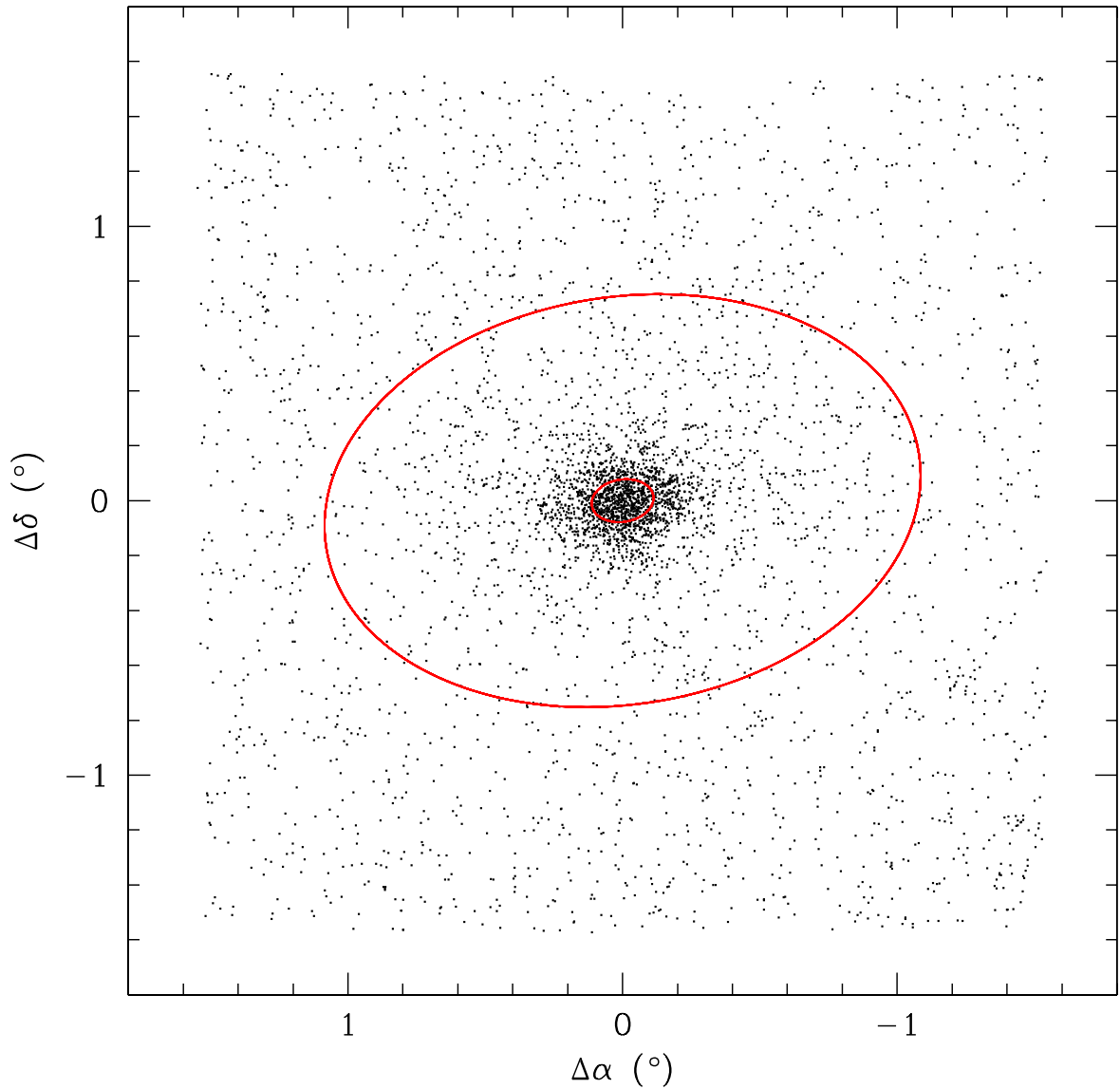}
\figcaption[Spatial distribution of Scl RGB stars]{Distribution of CMD-selected stars in the field of Sculptor, where the selection range is outlined in Fig.\ \ref{sclcmdlines}.  The inner and outer ellipses are the core and tidal radii respectively from the best-fitting King model to the RGB-selected stars, shown in Fig.\ \ref{sclradial}; $r_c = 6.8 \pm 1.2$ arcmin, $r_t = 72.5 \pm 4.0$ arcmin.  \label{sclrgbxy}}
\end{figure}

\begin{figure}
\plotone{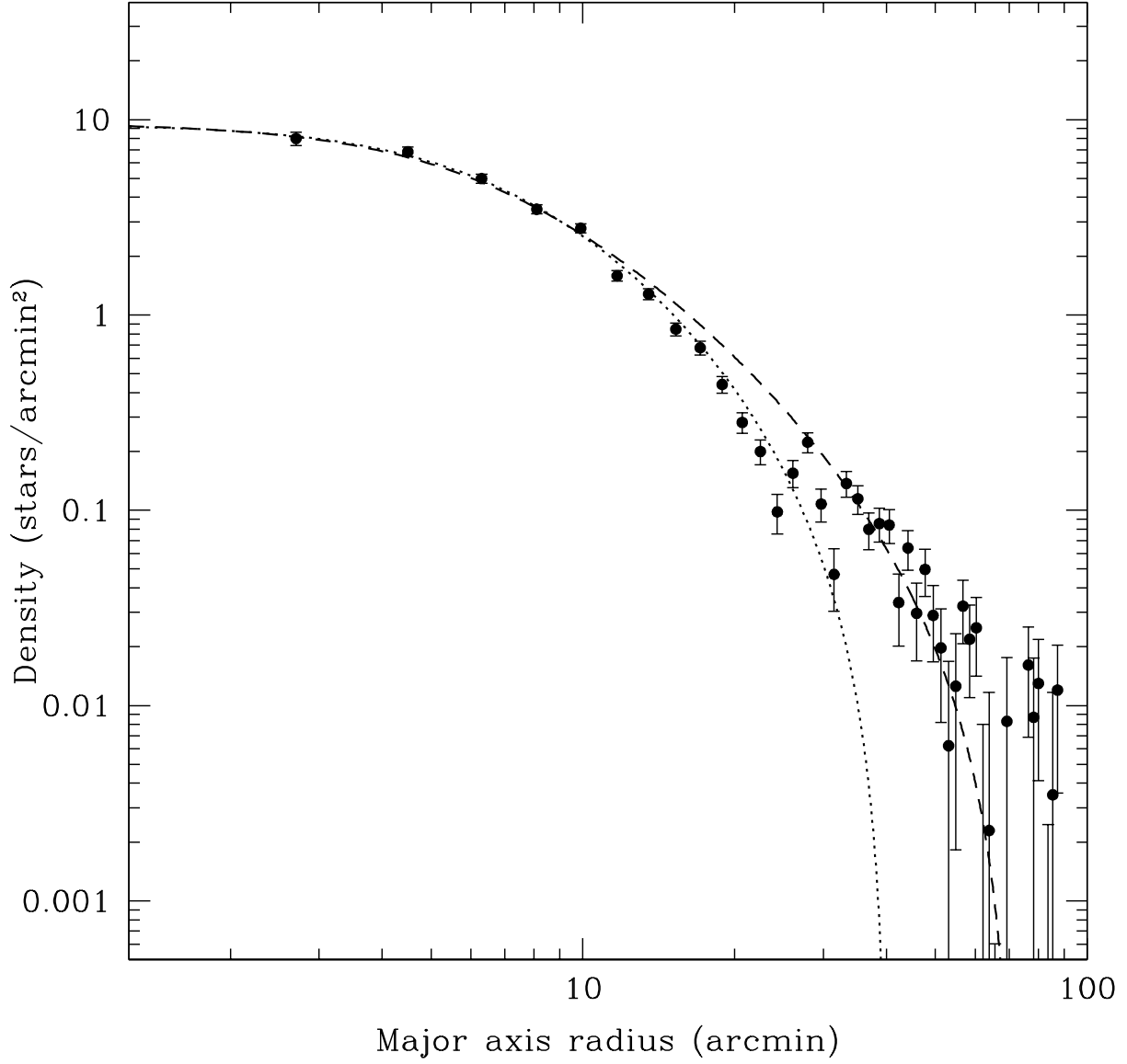}
\figcaption[Radial profile of Scl RGB stars]{Radial profile of the Scl RGB-selected stars, where the error bars are determined from Poisson noise.  A background level has been subtracted from all data points, calculated as $0.0893 \pm 0.0104$ stars/arcmin${}^2$ from all density values beyond a major axis radius of $80'$.  The dashed line represents the best-fitting King model to all points, and the dotted line traces the model fitted to all points within $40'$ of the Scl centre. \label{sclradial}}
\end{figure}

\begin{figure}
\plotone{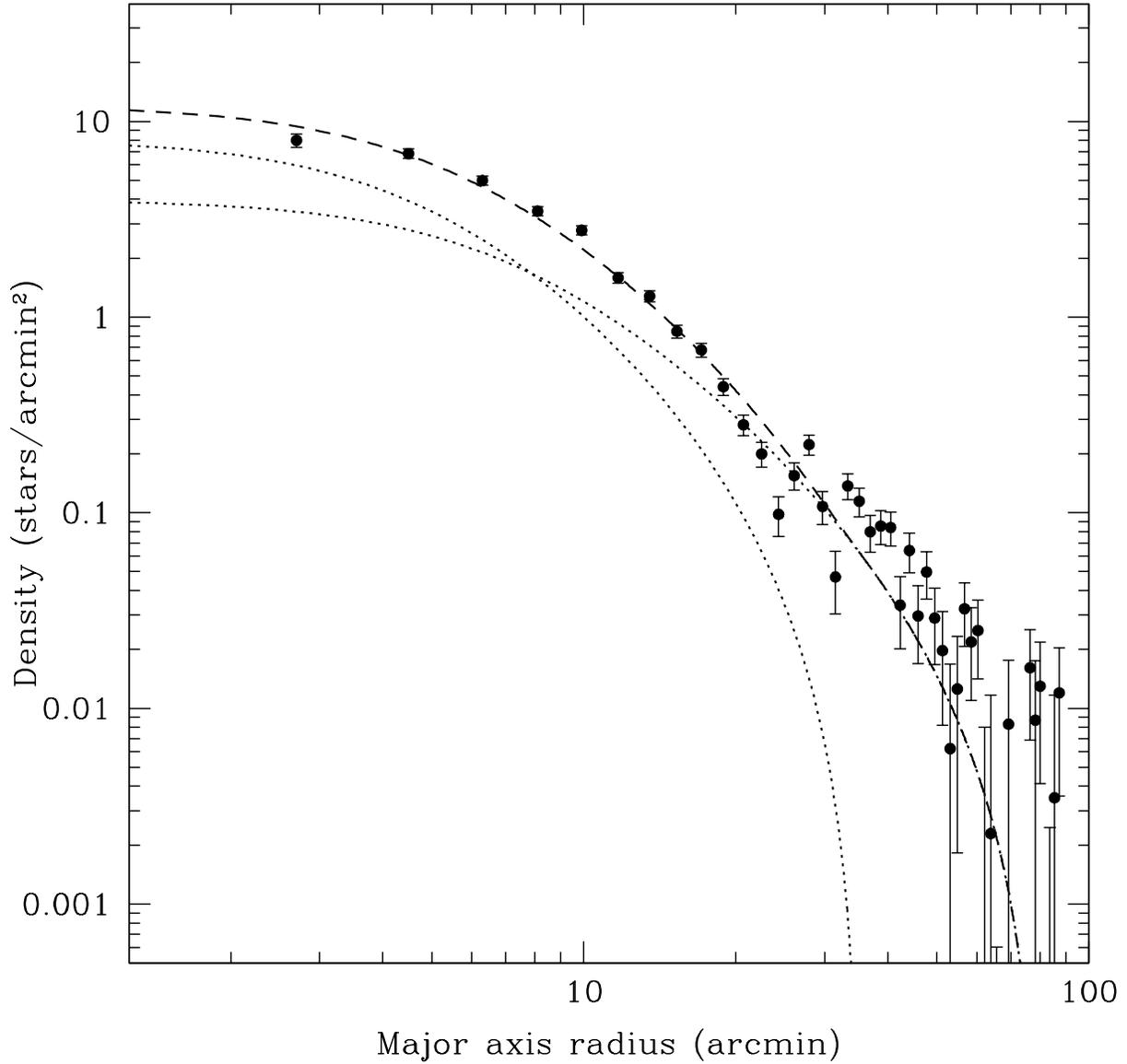}
\figcaption[Two component radial profile of Scl RGB stars]{Two component radial profile of the Scl RGB-selected stars.  The datapoints are the same as those in Fig.\ \ref{sclradial}, and the dashed line represents the two-component model.  Both components are shown as a dotted line, and their core and limiting radii, and their relative central densities, are listed as the RHB and BHB populations in Table \ref{kingpar}. \label{tworadial}}
\end{figure}

\begin{figure}
\plotone{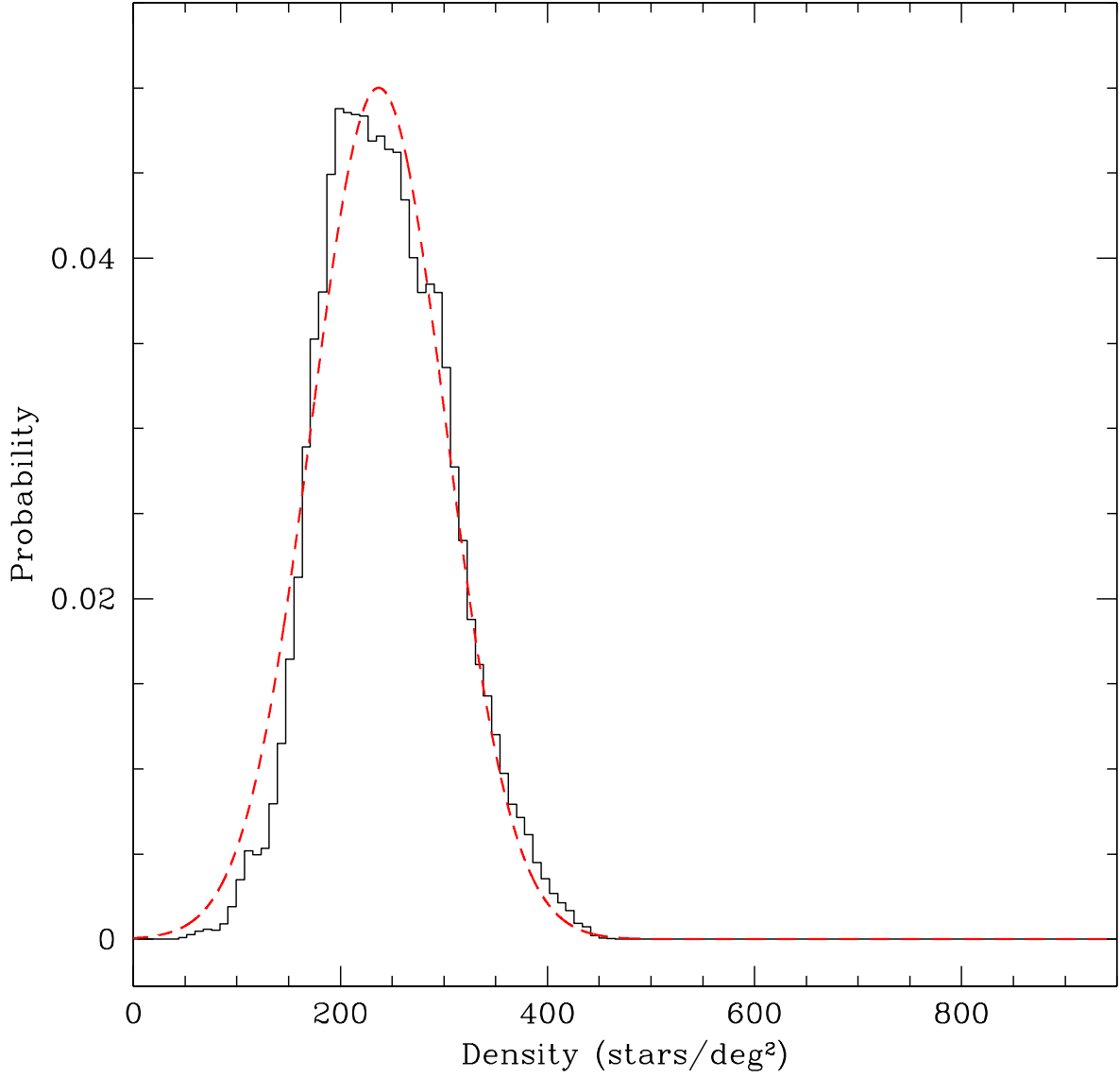}
\figcaption[Density Distribution]{Monte Carlo simulation of the density distribution of Scl RGB-selected stars beyond the nominal tidal radius, showing the probability of measuring a given density outside the nominal tidal radius.  The dashed line traces the best-fit Gaussian function to the data. \label{mc_xy}}
\end{figure}

\begin{figure}
\plotone{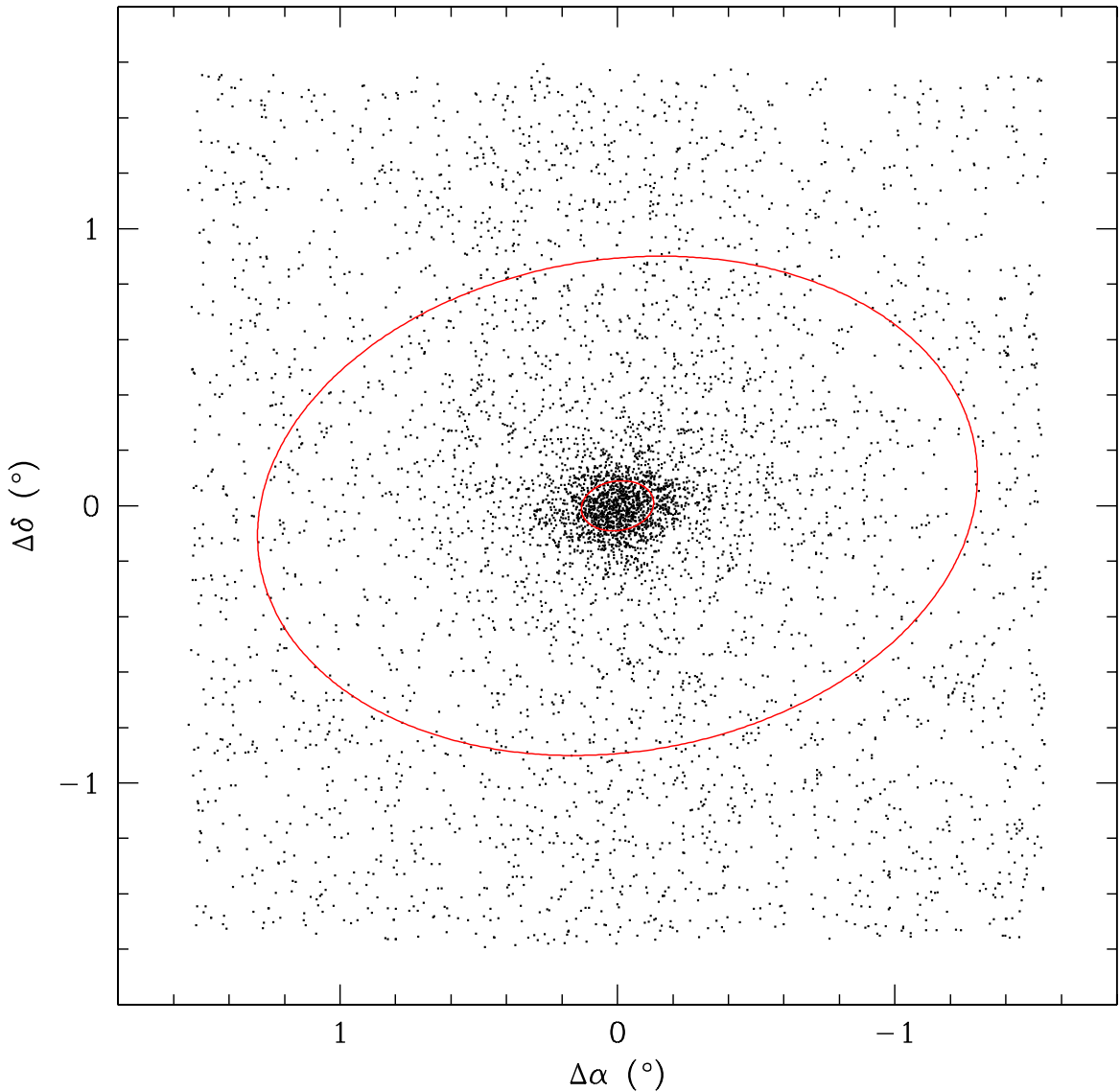}
\figcaption[Scl RGB dataset with superimposed $2.0\sigma$ artificial tidal tails]{The Scl RGB dataset, with artificial tidal tails extending in the North/South direction from the centre of the system.  The tails have a stellar density $2.0\sigma$ above that of the field population, and occupy uniformly the region $-0.5^{\circ} \le \Delta \alpha \le 0.5^{\circ}$ over the declination range of the survey, both inside and outside the nominal tidal radius. \label{sclrgbtails}}
\end{figure}

\begin{figure}
\plotone{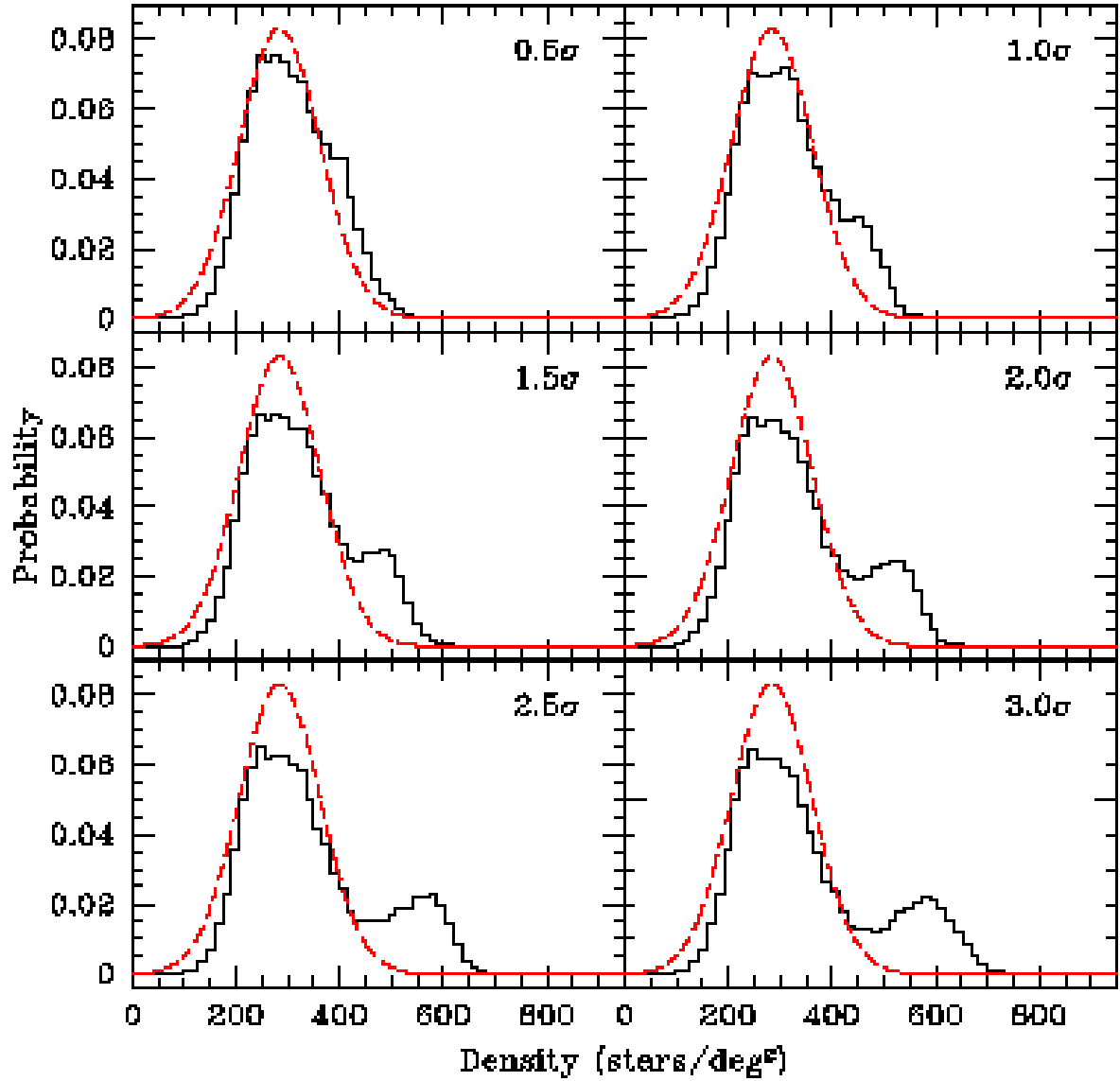}
\figcaption[Density probability functions measured for the six artificial tidal tail datasets]{The density probability functions for simulated tidal tails, where the mean stellar density of the tails above the background is given in the upper right corner of each panel.  \label{mc_xy_tot}}
\end{figure}

\begin{figure}
\plotone{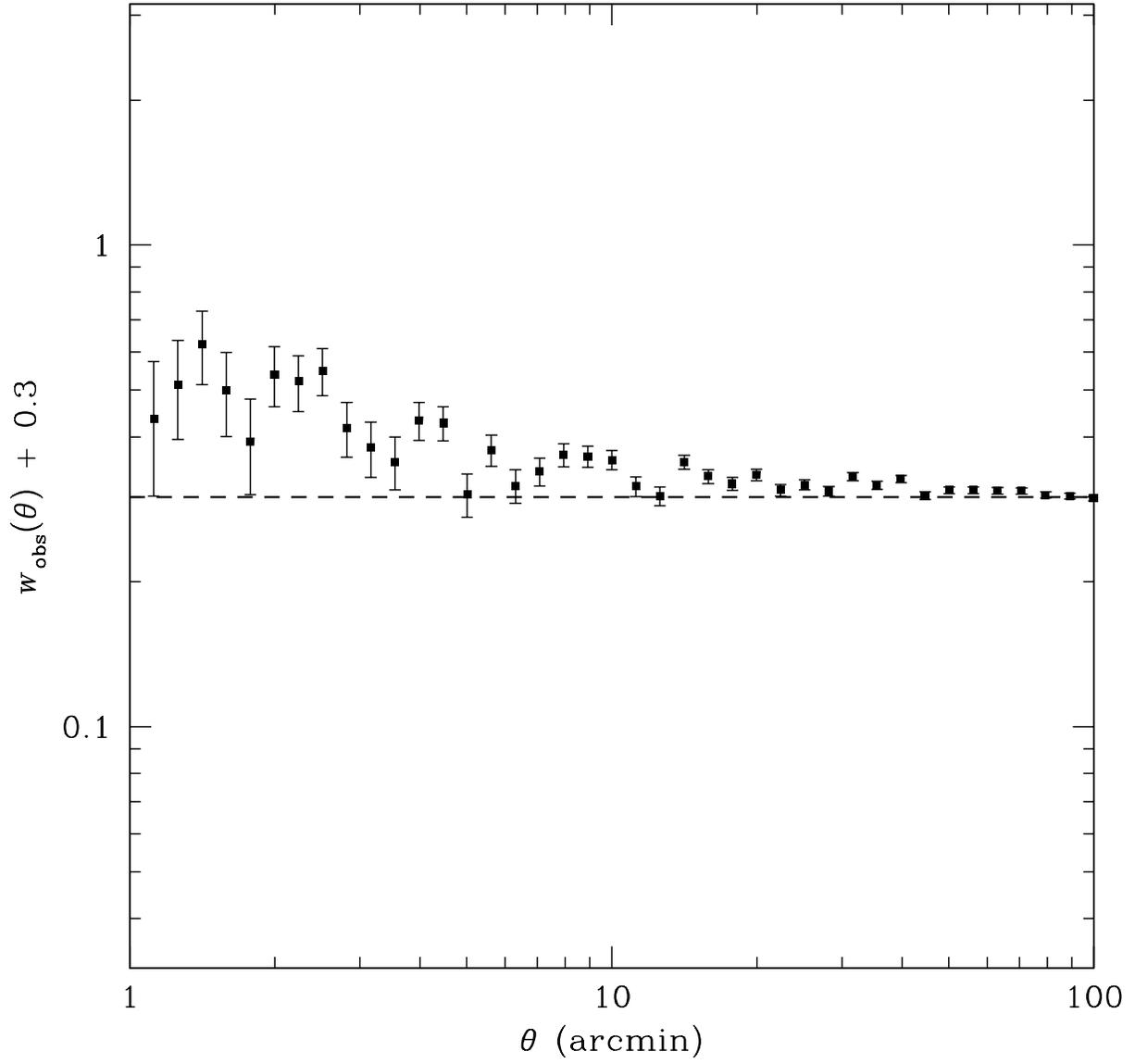}
\figcaption[Angular Correlation Function]{The {\em observed} angular correlation function for the extra-tidal region of the Sculptor dataset.  Error bars are measured from Poisson noise, and  factor of 0.3 has been added to all points.  The dashed line represents $w_{\mbox{\scriptsize obs}}(\theta) = 0$. \label{corrfn}}
\end{figure}

\begin{figure}
\plotone{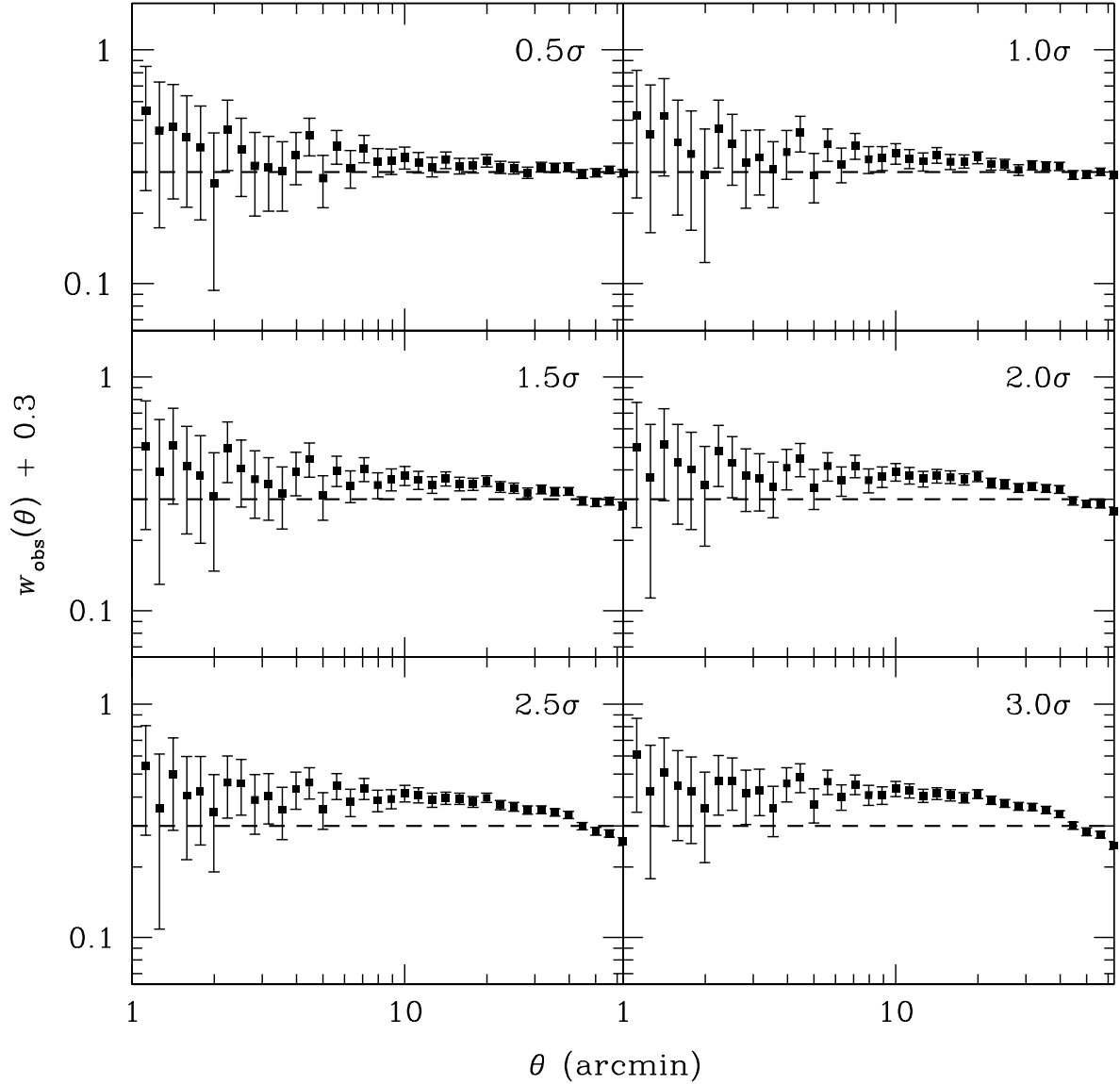}
\figcaption[Angular Correlation Function for the tidal tail simulations]{The observed angular correlation function measured for the simulated tidal tails, where the overdensity of the tails is indicated in the upper right corner of each panel.  Error bars are measured from Poisson noise, and a factor of 0.3 has been added to all points.  The dashed line in each panel represents $w_{\mbox{\scriptsize obs}}(\theta) = 0$, the expected angular correlation function for a normal distribution of stars.  \label{acftest}}
\end{figure}

\begin{figure}
\plotone{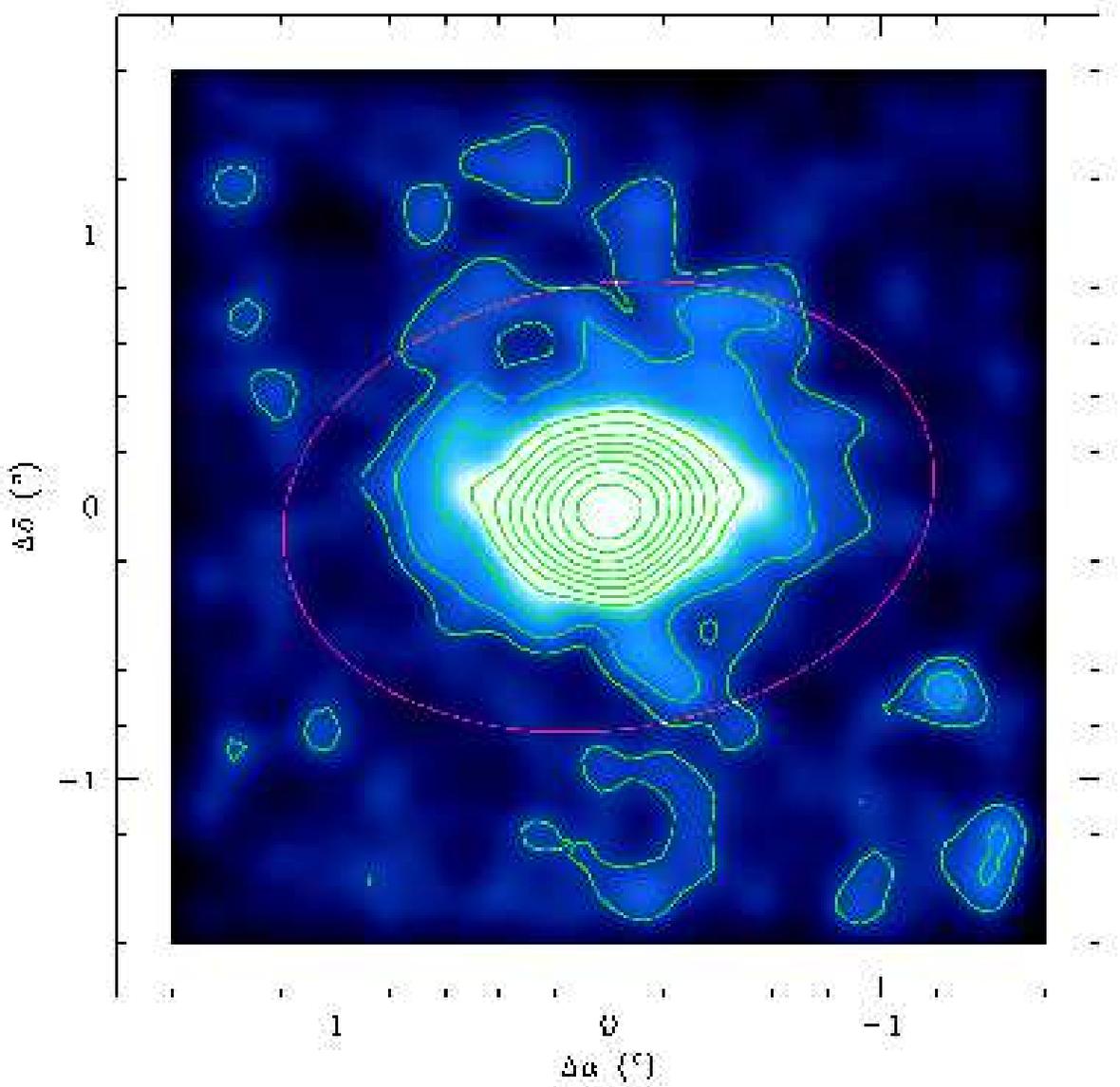}
\figcaption[Sculptor Contour Plot]{Distribution of Sculptor RGB stars ($V \le 20$) where each star has been convolved with a Gaussian of radius $40''$.  The contours are logarithmically spaced, and are fitted with a smoothing length of $3.0'$.  The first and second contours represent stellar densities $1\sigma$ and $3\sigma$ above the field star population respectively.  The red line represents the nominal tidal radius of Sculptor. \label{sclcontour}}
\end{figure}

\begin{figure}
\plotone{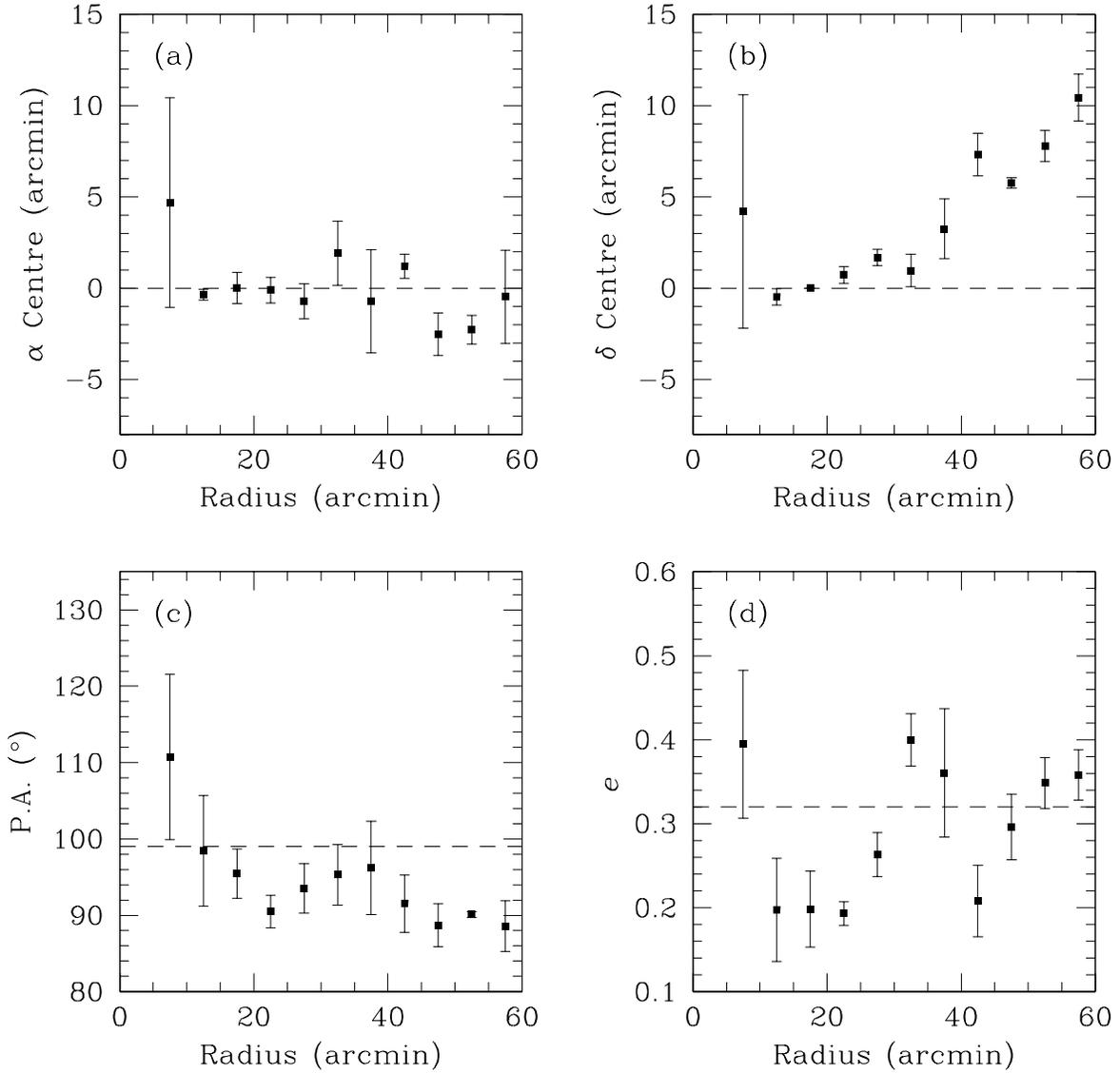}
\figcaption[Radial Dependence of Structure]{The dependence of Scl structure on radius, where each point represents the binned result from five measurements at succeeding radii.  The dashed lines indicate the values listed in \citet{m98}.  Error bars are the standard deviations of each bin. \label{lscontour}}
\end{figure}

\begin{figure}
\plotone{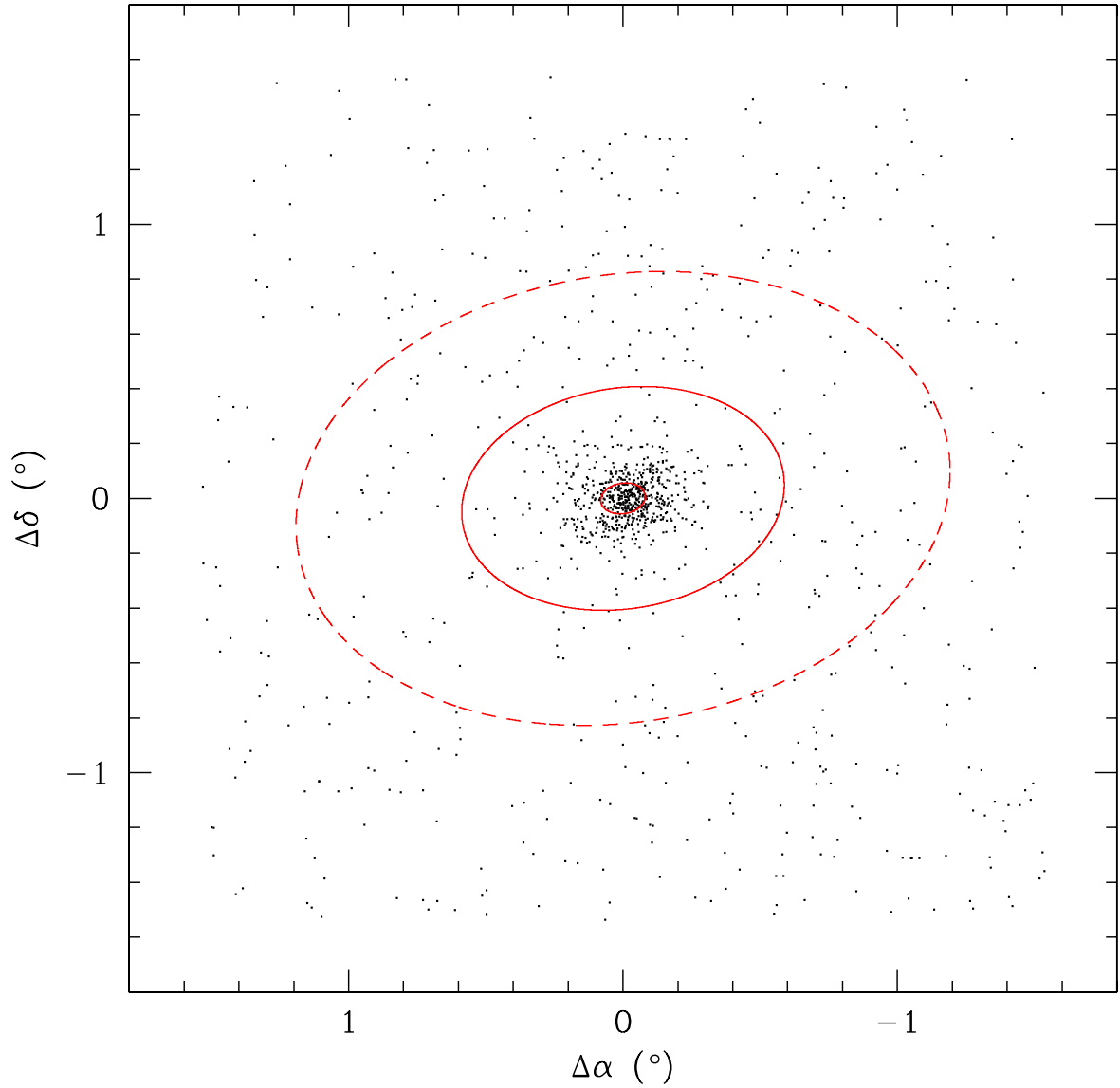}
\figcaption[Spatial distribution of Scl RHB stars]{Distribution of Scl RHB-selected stars.  The inner and outer ellipses represent the core and limiting radii of the RHB stars using the King profile shown in Fig.\ \ref{radialhb}(a) after removing the background.  The outer dashed line represents the limiting radius of the RGB stars.  \label{sclrhbxy}}
\end{figure}

To investigate the radial dependence of the HB morphology, we selected stars using the colour-magnitude ranges outlined in Fig.\ \ref{sclcmd_cand}.  The selection range for both components of the horizontal branch was chosen to minimise contamination effects due to stars in the RGB and instability strip.  First we examine the red HB stars.  The dashed line represents the completeness limit ($V=20.4$ and $I=19.6$) of the four inner fields.  Hence, the corresponding spatial distribution of RHB stars displayed in Fig.\ \ref{sclrhbxy} is essentially complete for all values of $\Delta \alpha$ and $\Delta \delta$ between $-0.6$ and $0.6$ deg.  A radial profile of the RHB candidate stars was assembled by selecting stars within a set of annuli of constant centre, ellipticity and position angle (listed by \citealt{m98}). The background stellar density was calculated as the mean value beyond $r=40'$ (Table \ref{kingpar}).  This value was subtracted from the stellar density measurements, and the resulting radial profile is shown in Fig.\ \ref{radialhb}(a), where the dashed line represents the best-fitting King profile.  The resulting core and limiting radii are represented by the solid ellipses in the RHB-selected spatial map (Fig.\ \ref{sclrhbxy}), while the dashed line represents the limiting radius of the Scl RGB stars (see Table \ref{kingpar}).

Now we turn to the blue HB stars.  The BHB selection box shown in Fig.\ \ref{sclcmd_cand} lies beyond the photometric limit of the four inner fields.  Hence, to determine which fields contain a near-complete sample at this luminosity, we placed $\sim$5000 artificial BHB stars in each field and attempted to recover these using the same photometry routines described in \S \ref{photsurvey} above.  The artifical stars were clustered around $V=20.35$, $(V-I) = 0.3$ with a dispersion of $0.15$ magnitudes in both $V$ and colour, resembling the appearance of the BHB as seen in Fig.\ \ref{sclcmd_cand}.  The recovery statistics for each field are listed in Table \ref{bhbstats}.  For each artificial star, we measured the difference between the input and output magnitudes.  The values $\sigma_V$ and $\sigma_I$ are the standard deviations from the mean difference, and hence indicate the level of photometric accuracy achieved by each field at the depth of the BHB.

\begin{figure}
\plotone{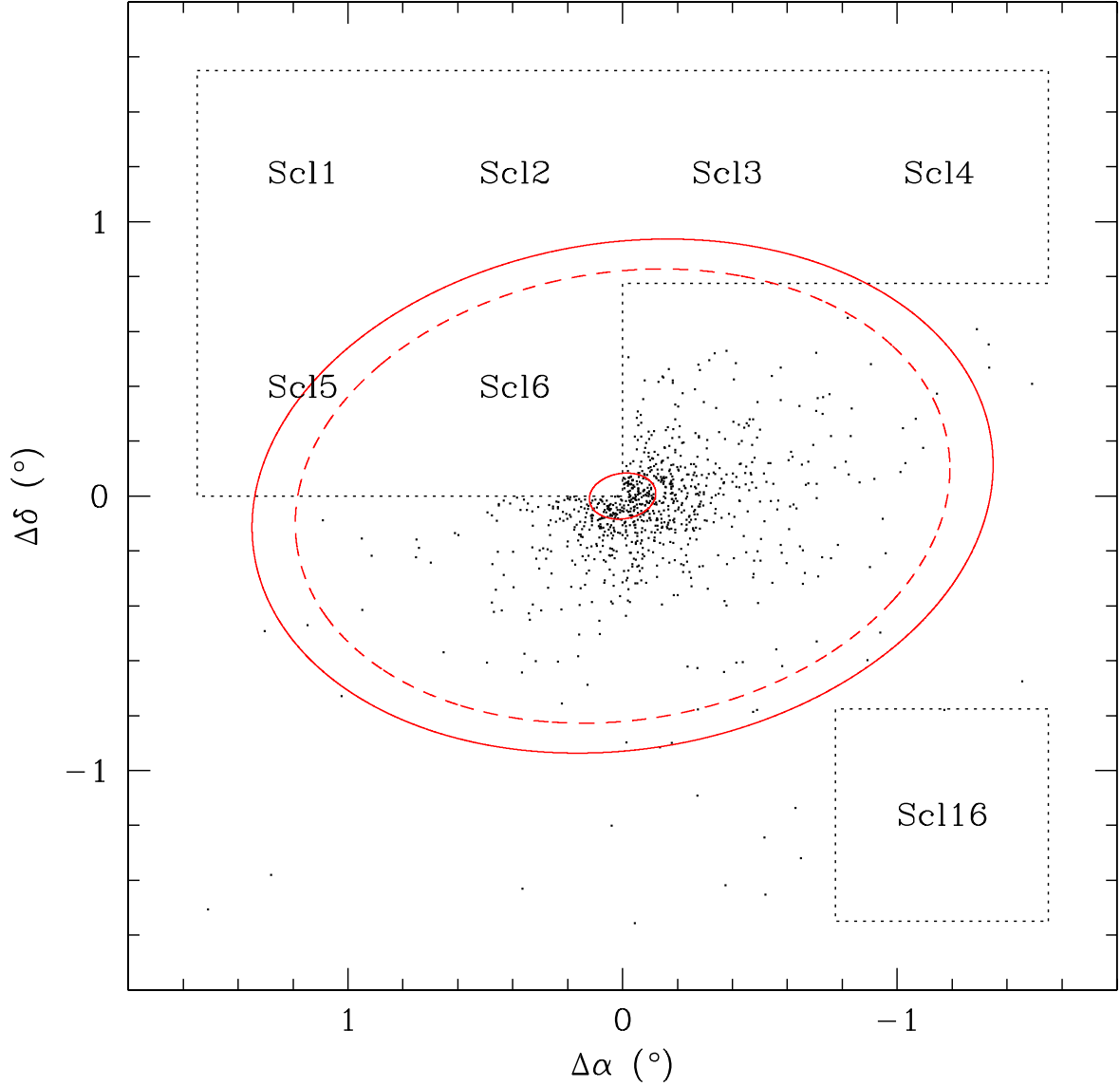}
\figcaption[Spatial distribution of Scl BHB stars]{Distribution of Scl BHB-selected stars, where the outlined fields have been removed due to completeness effects.  The inner and outer ellipses represent the core and limiting radii of the BHB stars using the King profile shown in Fig.\ \ref{radialhb}(b).  The dashed line represents the limiting radius of the RGB stars.  \label{sclbhbxy}}
\end{figure}

\begin{figure}
\plotone{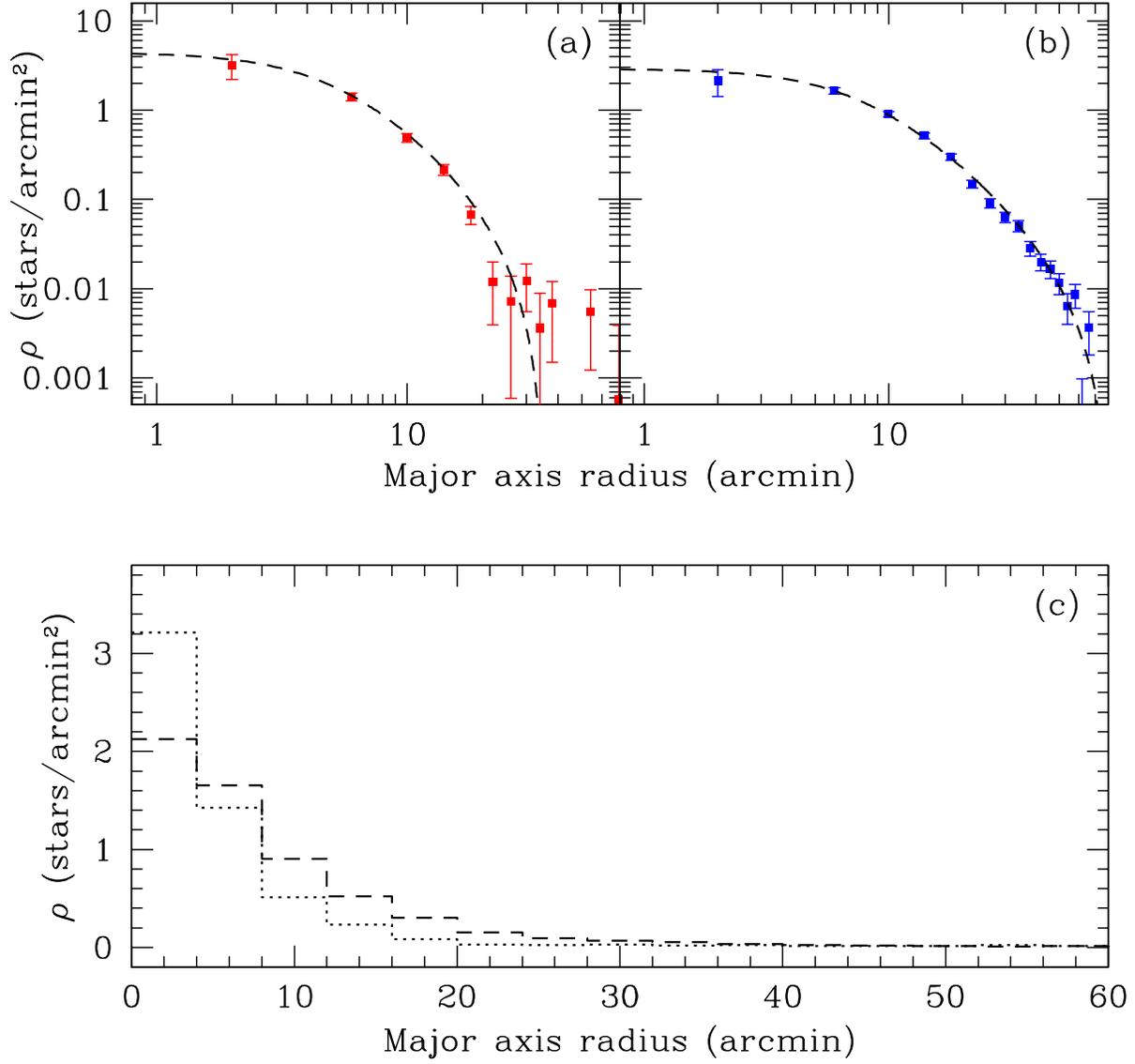}
\figcaption[Radial profile of HB stars]{(a) Radial profile of the RHB stars.  A background level has been subtracted from all data points (see text).  The errors are calculated from Poisson noise, and the dashed line represents the best-fitting King model.  (b) Same as (a) for the BHB stars.  (c) Radial distribution of the blue (dashed line) and red HB (dotted line) stars.  No background correction has been made for these histograms. \label{radialhb}}
\end{figure}

An examination of Table \ref{bhbstats} reveals that seven of the sixteen fields recovered significantly less than $80\%$ of the artificial BHB stars.  Therefore, BHB stars were selected only in the remaining nine fields, and their spatial distribution is shown in Fig.\ \ref{sclbhbxy}.  The dotted line outlines the seven incomplete fields, which are also labelled.  We constrained the best-fitting BHB radial profile from the nine complete fields using the same technique described for the RHB population above.  Fig.\ \ref{radialhb}(b) shows the BHB profile, where the errors are the Poission noise for each annulus and the dashed line represents the best-fitting King model.  We subtracted a background density from the profile, measured to be $(2.97 \pm 0.20) \times 10^{-3}$ stars/arcmin${}^2$ in the area beyond a major axis radius of $80'$.  The resulting King profile parameters are listed in Table \ref{kingpar}, while the core and limiting radii are displayed in the BHB-selected spatial map (Fig.\ \ref{sclbhbxy}).  A comparison of the radial profiles for the red and blue horizontal branch stars is shown in Fig.\ \ref{radialhb}(c).  We have thus confirmed the HB gradient in Scl, and measured the spatial extent of the two populations described by \citet{tolstoy04}.

\section{Results -- Spectroscopic Survey} \label{specresults}
The Scl spectroscopic dataset contained 723 candidate RGB stars with radial velocity errors less than 20 km s${}^{-1}$.  The velocity distribution of these stars is shown in Fig.\ \ref{sclspecV}.  The population is clearly bimodal in velocity, with the Scl population clustered around $v_r \sim 100$ km s${}^{-1}$ and the field population around $\sim$0 km s${}^{-1}$.  To determine the mean Scl radial velocity, we selected 188 stars in the radial velocity range 76.5 km s${}^{-1}$ $\le v_r \le 137.5$ km s${}^{-1}$, and calculated the mean velocity using the relation from \citet{arman86},
\begin{equation} \label{meanvel}
\overline{v} = \sum_{i=1}^{N}{w_i v_i} {\Big \slash} \sum_{i=1}^{N}{w_i},
\end{equation}
\noindent where the weight ($w_i$) was defined to be the inverse square of the individual velocity uncertainty for each velocity $v_i$.  The mean heliocentric velocity of Scl was then measured to be,
\begin{displaymath}
\overline{v}_{\mbox{\scriptsize Scl}} =  106.9 \pm 1.1 {\mbox{ km s${}^{-1}$}}
\end{displaymath}
\noindent where the error is the weighted standard deviation of the mean.  This value is marked as the dashed line in Fig.\ \ref{sclspecV}, and agrees well with the previous measurements by \citet{arman86} and \citet{queloz95}.

\subsection{Sculptor Membership}
In this section we describe several methods used to eliminate foreground Galactic stars from the Scl member list.  The primary method was provided by a radial velocity selection range.  We measured the velocity dispersion around the Scl mean to be $\sigma_{\mbox{\scriptsize obs}} = 15.2 \pm 0.6$ km s${}^{-1}$.  Assuming the average global velocity dispersion of the Scl system is approximately 10 km s${}^{-1}$ \citep{tolstoy04}, then this $\sigma_{\mbox{\scriptsize obs}}$ value agrees with the overall velocity uncertainty of $\sim$10 km s${}^{-1}$.  This allows us to define the velocity selection range for the Scl members.

\subsubsection{Velocity Selection and the Field Population}
\begin{figure}
\plotone{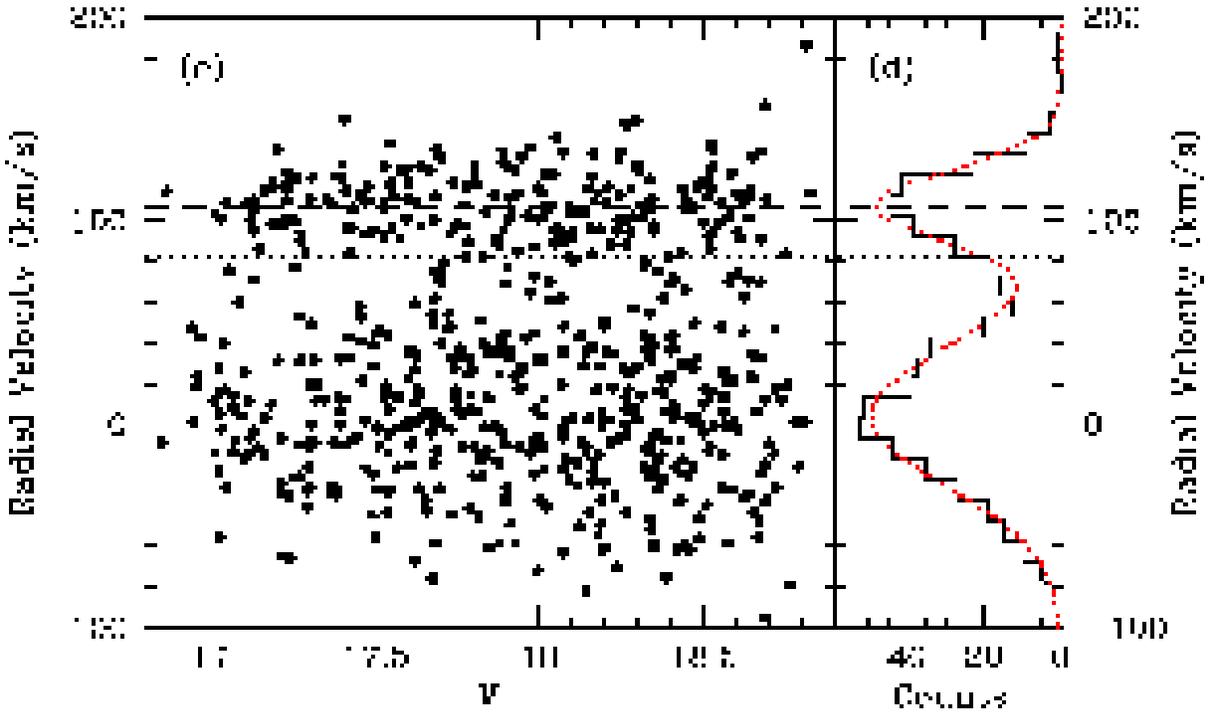}
\figcaption[Scl survey radial velocities]{(a) Radial velocities of all candidate RGB stars against $V$ magnitude, where the dashed line illustrates the mean Scl velocity.  The dotted lines represent the $2.5\sigma_{\mbox{\scriptsize obs}}$ selection range for candidate Scl stars.  (b) Histogram for the velocities in (a).  The dotted line represents the best-fitting bimodal Gaussian function. \label{sclspecV}}
\end{figure}

The dotted lines in Fig.\ \ref{sclspecV} illustrate the velocity range for the selection of Scl stars.  We have chosen a width of $2.5\sigma_{\mbox{\scriptsize obs}}$ to either side of the mean Scl velocity ($81.9 \le v_r \le 131.9$ km s${}^{-1}$).  This velocity range gave 157 likely members of the Scl system.  An important concern in this selection criterion is residual background noise from Galactic stars; Fig.\ \ref{sclspecV} shows the distribution in velocity of these foreground stars included in the Scl observations, and clearly some portion of the Galaxy's population extends into the Scl velocity range.  The colour selection range for these stars illustrated in Fig.\ \ref{sclcmd_cand} allows two possible populations of stars to contribute to the field population; halo RGB stars or Galactic (thin and/or thick) disk main sequence stars.  Assuming the field population comprises giant stars, then the apparent magnitude selection range implies a distance of at least 45 kpc (from the Yonsei-Yale isochrones; \citealt{yi01,kim02}).  The stellar halo contains $\sim$$1\%$ of the stellar mass of the Galaxy, and has a power density distribution $\rho \propto r^{-n}$ where $n = 2.5 - 3.5$ depending on the stellar population \citep{chiba00,chen01,vivas01}, thus we expect the contribution of halo stars to the field population to be negligible.

In contrast, if we assume the field population comprises dwarf stars occupying the Galactic disk, then the colour-magnitude selection range imply these stars are located at a distance of $350 \le d \le 1600$ pc.  Now, the thin disk displays a scale height of $\sim$300 pc (for example, \citealt{gilmore84,bahcall84,chen01}), whereas the scale height of the thick disk \citep{gilmore83} is approximately 1000 pc (\citealt{chen01}; see \citealt{norris99} for a recent review).  Therefore, we expect a negligible contribution from the thin disk, and hence the field population composed almost entirely of Galactic thick disk stars.

The Scl dSph is located near to the South Galactic Pole (SGP), hence the width in velocity space of the field stars is determined almost exclusively by the vertical velocity dispersion of the Galactic thick disk.  Recent measurements \citep{norris87,layden96,chiba00} have found this value to be $30 - 35$ km s${}^{-1}$.  We calculated the {\em observed} velocity dispersion of the field stars towards Scl to be 34.1 km s${}^{-1}$.  Assuming an instrumental uncertainty of 10 km s${}^{-1}$ gave the true velocity dispersion of these stars to be 32.6 km s${}^{-1}$ which agrees with previously determined values.  There were 364 stars in the velocity range $-50 \le v_r \le 50$ km s${}^{-1}$, hence we deduce the dataset includes approximately 430 Galactic stars overall.  Assuming the velocity distribution follows the Gaussian function illustrated in Fig.\ \ref{sclspecV}(b), then the $2.5\sigma_{\mbox{\scriptsize obs}}$ star selection criterion is expected to include approximately eight high-velocity Galactic thick disk stars.

\subsubsection{Equivalent Widths}
\begin{figure}
\plotone{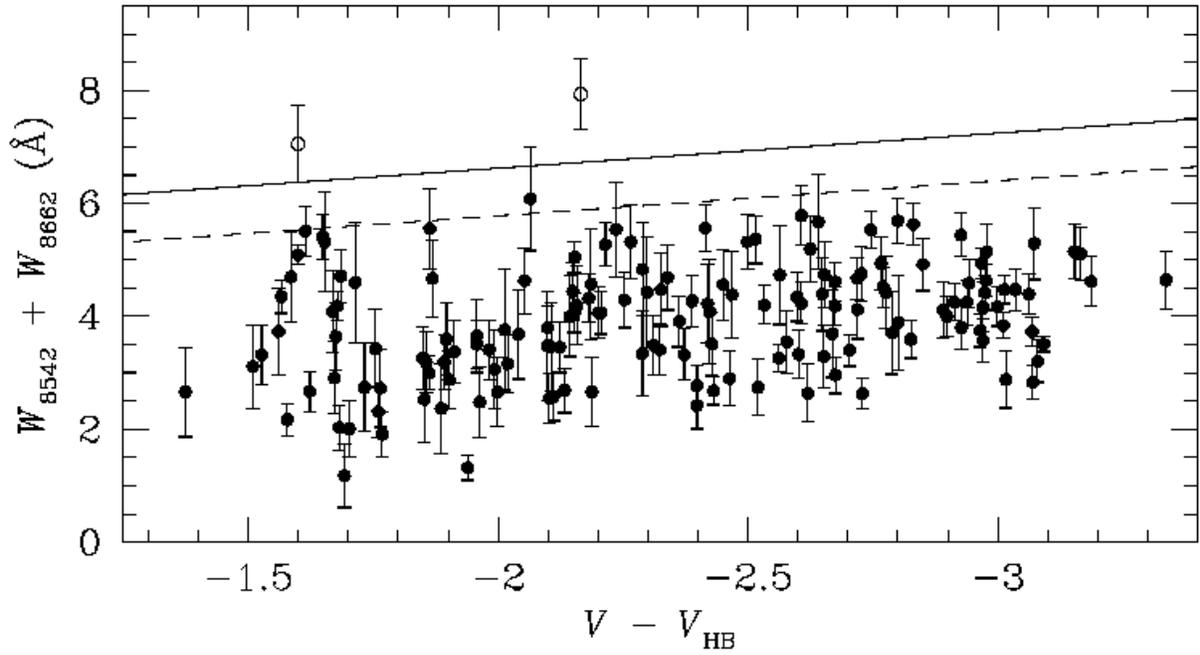}
\figcaption[Ca {\sc ii} line strengths for Sculptor stars]{Total Ca {\sc ii} line strengths for the candidate Scl stars selected using the velocity range in Fig.\ \ref{sclspecV}.  All stars with an equivalent width uncertainty greater than $3\sigma_W$ have been removed, and the open circles represent stars outside the abundance selection range.  Errorbars are from Gaussian fitting uncertainties.  The dashed line traces expected abundance limit of Scl ([Fe/H] $\le -1.0$; \citealt{tolstoy04}).  The solid lines represents this limit vertically shifted by $2\sigma_W = 0.84$ \AA.  The two stars above this limit rejected as members are represented by open circles. \label{eqwidthobj}}
\end{figure}

A second test for the membership status of the candidate Scl stars was provided by abundance measurements.  From the psuedo-equivalent widths we were able to calculate $\Sigma \mbox{Ca}$ for the 157 candidate Scl stars, which can be directly related to [Fe/H] for red giants.  Given that any residual field objects would be foreground dwarf stars, we expect their surface gravities may provide a Ca triplet measurement markedly different from the range seen in Sculptor.  The resulting $\Sigma \mbox{Ca}$ values are plotted against magnitude in Fig.\ \ref{eqwidthobj}.  Based on Ca {\sc ii} triplet observations of approximately 300 RGB stars in Scl, \citet{tolstoy04} observed an abundance range in Scl of $-2.8 \lesssim$ [Fe/H] $\lesssim -0.9$.  The upper abundance limit is marked as the dashed line in Fig.\ \ref{eqwidthobj}.  However, the line strength index has an uncertainty of $\sigma_W = 0.42$ \AA, therefore we extended the $W'$ selection limit by a factor of $2\sigma_W$, and this limit is marked by the solid line in Fig.\ \ref{eqwidthobj}.  In order for a star to be included as a Scl member, its line strength must lie below this value.  The open circles represent the two stars which do not fulfill this criterion and have been discarded as Scl members.

\begin{figure}
\plotone{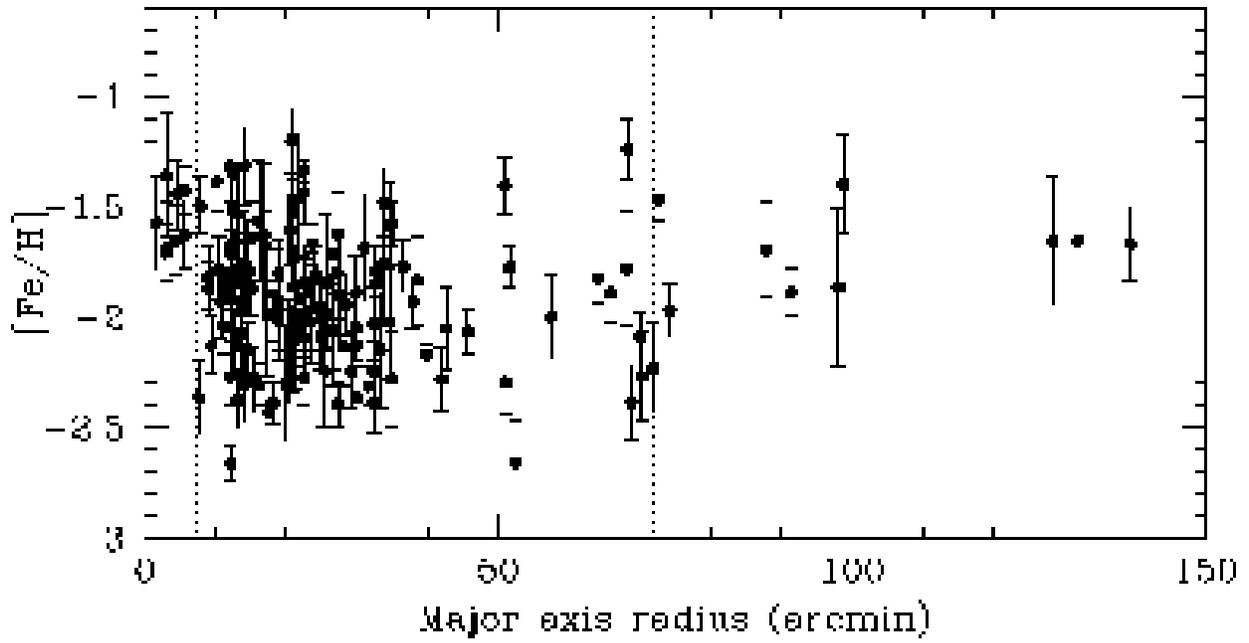}
\figcaption[Metallicity vs radius for the upper RGB Scl stars]{Metallicity against radius for the 148 Scl members.  [Fe/H] has been calculated from the Ca {\sc ii} equivalent widths.  The dotted lines represent the core and tidal radii of the Scl RGB population.  \label{radmet}}
\end{figure}

\citet{tolstoy04} revealed that Scl contains a metallicity gradient, such that the mean [Fe/H] value of the inner region is higher than that of the outer region.  We found a similar result in our data.  Fig.\ \ref{radmet} shows the metallicity of these stars as a function of (major axis) radius, under the assumption that Scl has an ellipticity of $e=0.32$ \citep{m98}.  The two dashed lines mark the core and nominal tidal radii.  We subdivided these data at the core radius ($r_c$), and found the mean metallicity for both datasets to be,
\begin{displaymath}
\langle \mbox{[Fe/H]} \rangle = \left \{
\begin{array}{lll}
-1.56 \pm 0.04 & \mbox{~if ~} r \le r_c; & N=8  \\
-1.92 \pm 0.03 & \mbox{~if ~} r_c < r < r_t; & N=135  \\
\end{array} \right.
\end{displaymath}
\noindent where the errors are the standard deviation of the mean.  That is, we have confirmed the metallicity gradient noted by \citet{tolstoy04}.

\subsubsection{Proper Motions}
\begin{figure}
\plotone{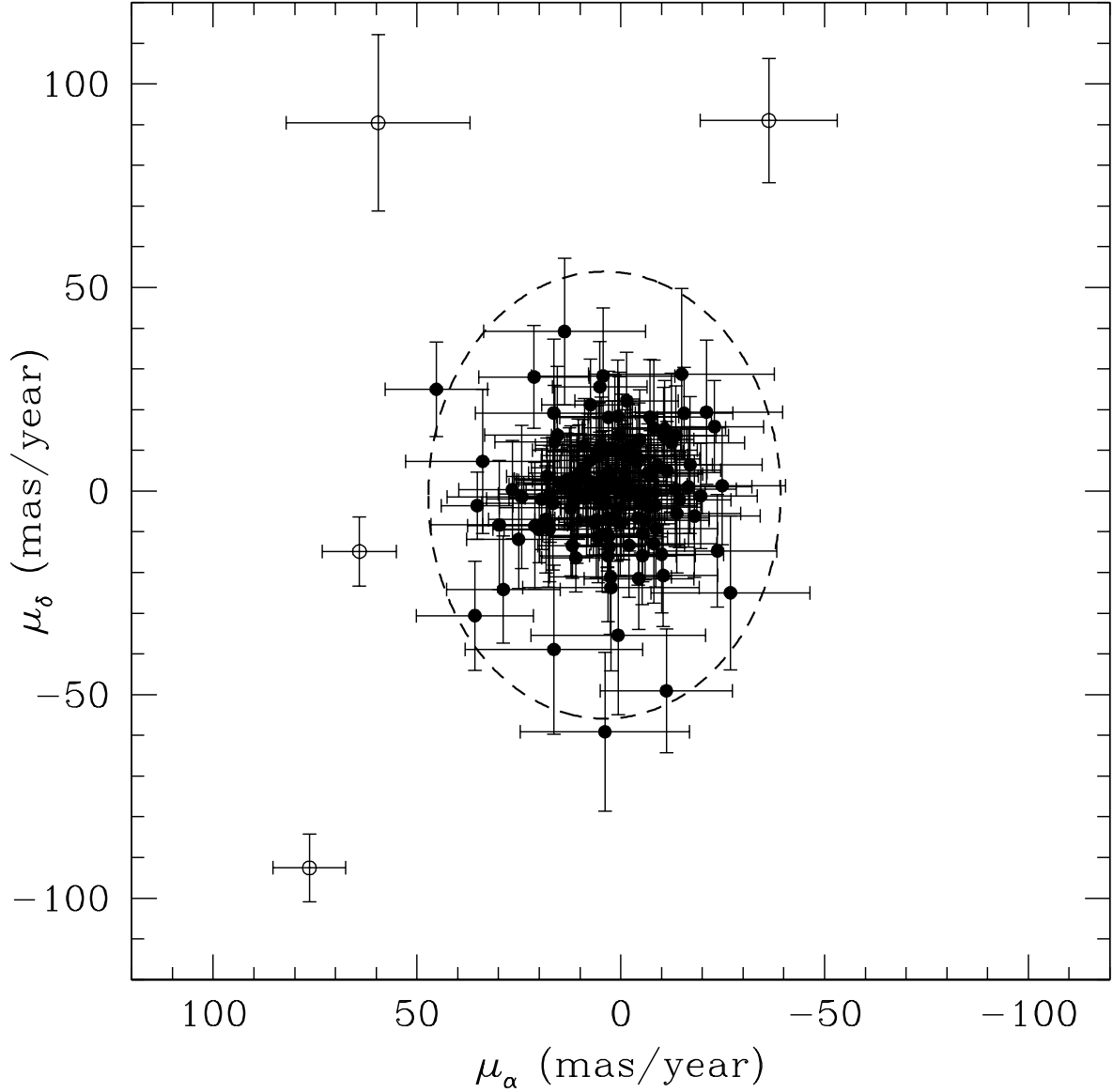}
\figcaption[Proper motion of candidate Scl stars]{Proper motions (in milliarcsec/year) of all stars within the radial velocity (Fig.\ \ref{sclspecV}) and Ca {\sc ii} line strength (Fig.\ \ref{eqwidthobj}) ranges.  The dashed ellipse outlines the selection range for Scl stars, and has a minor (major) axis radius of $3\sigma_{\mu,\alpha}$ ($3\sigma_{\mu,\delta}$).  The parameters $\sigma_{\mu,\alpha}$ and $\sigma_{\mu,\delta}$ are the standard deviations from the mean proper motion values $\overline{\mu}_{\alpha}$ and $\overline{\mu}_{\delta}$ respectively.  Stars represented by open circles are high proper motion objects rejected as Scl members. \label{propmot}}
\end{figure}

Scl is located towards the SGP, hence we expect many of the field stars to display a high proper motion.  Proper motions of all stars in the refined Scl member list were extracted from the SuperCOSMOS Science Archive\footnote{http://surveys.roe.ac.uk/ssa/index.html}, which makes use of observations at up to four different epochs.  The astrometry from this catalogue is described by \citet{hambly01}.  The proper motions of the potential members are shown in Fig.\ \ref{propmot}.  We measured the standard deviations from the mean motions in RA and Dec to be $\sigma_{\mu,\alpha} = 14.4$ milliarcsec/year and $\sigma_{\mu,\delta} = 18.3$ milliarcsec/year respectively.  Thus, the proper motion limits for Scl stars were set to be $3\sigma_{\mu,\alpha}$ and $3\sigma_{\mu,\delta}$, and this is marked as the dashed ellipse in Fig.\ \ref{propmot}.  The four stars with a proper motion (within their uncertainties) beyond this value were discarded as Scl members and are represented as open circles.

\subsection{Extra-Tidal Candidates}
\begin{figure}
\plotone{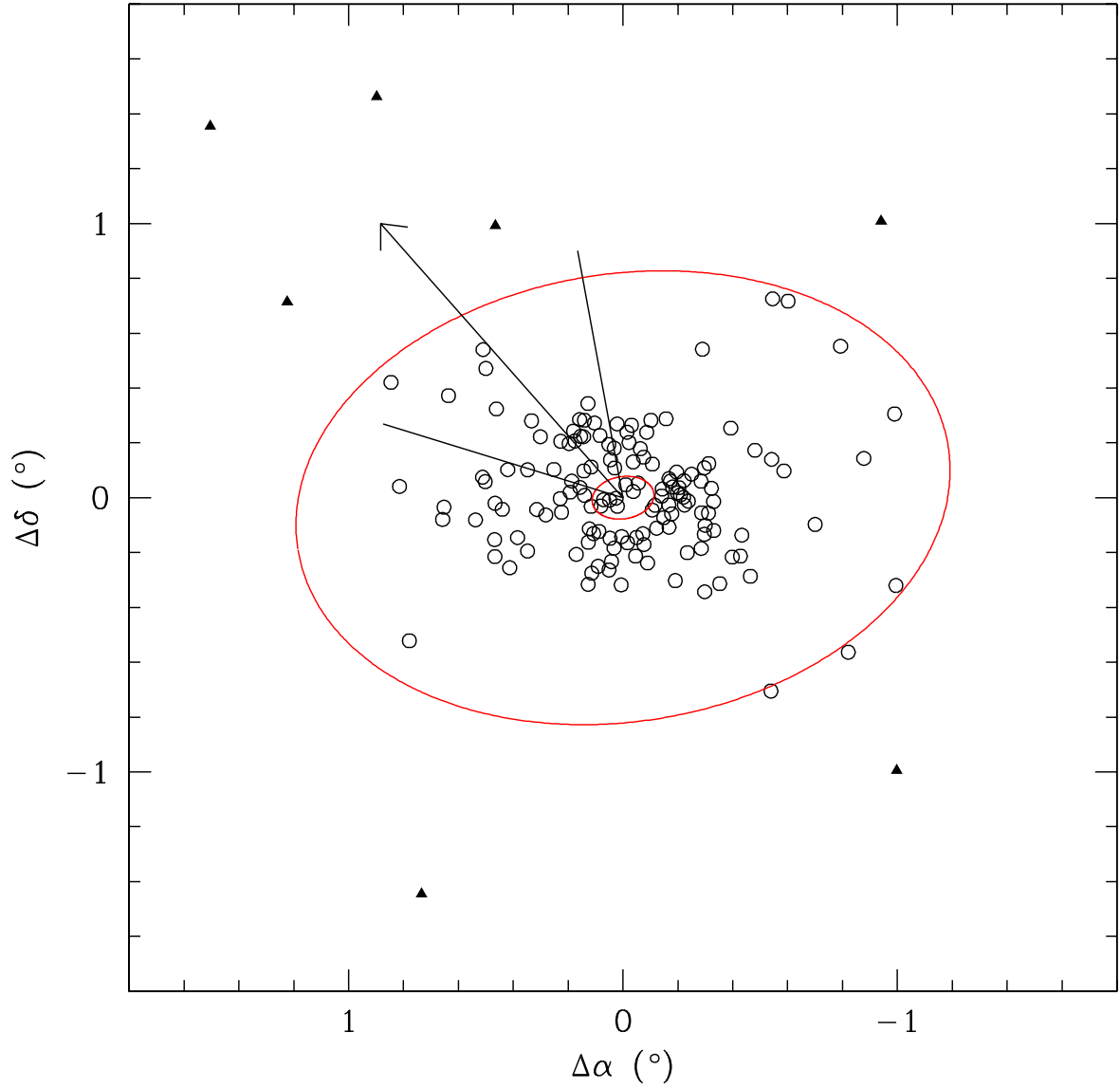}
\figcaption[Spatial distribution of Scl spectroscopic members]{Spatial distribution of the 148 Sculptor members.  The closed triangles represent those stars beyond the nominal tidal radius (including the tidal radius uncertainty).  The arrow indicates the proper motion measured by \citet{schweitzer95}, where the lines to either side represent the corresponding uncertainties.  \label{sclspecxy}}
\end{figure}

\begin{figure}
\plotone{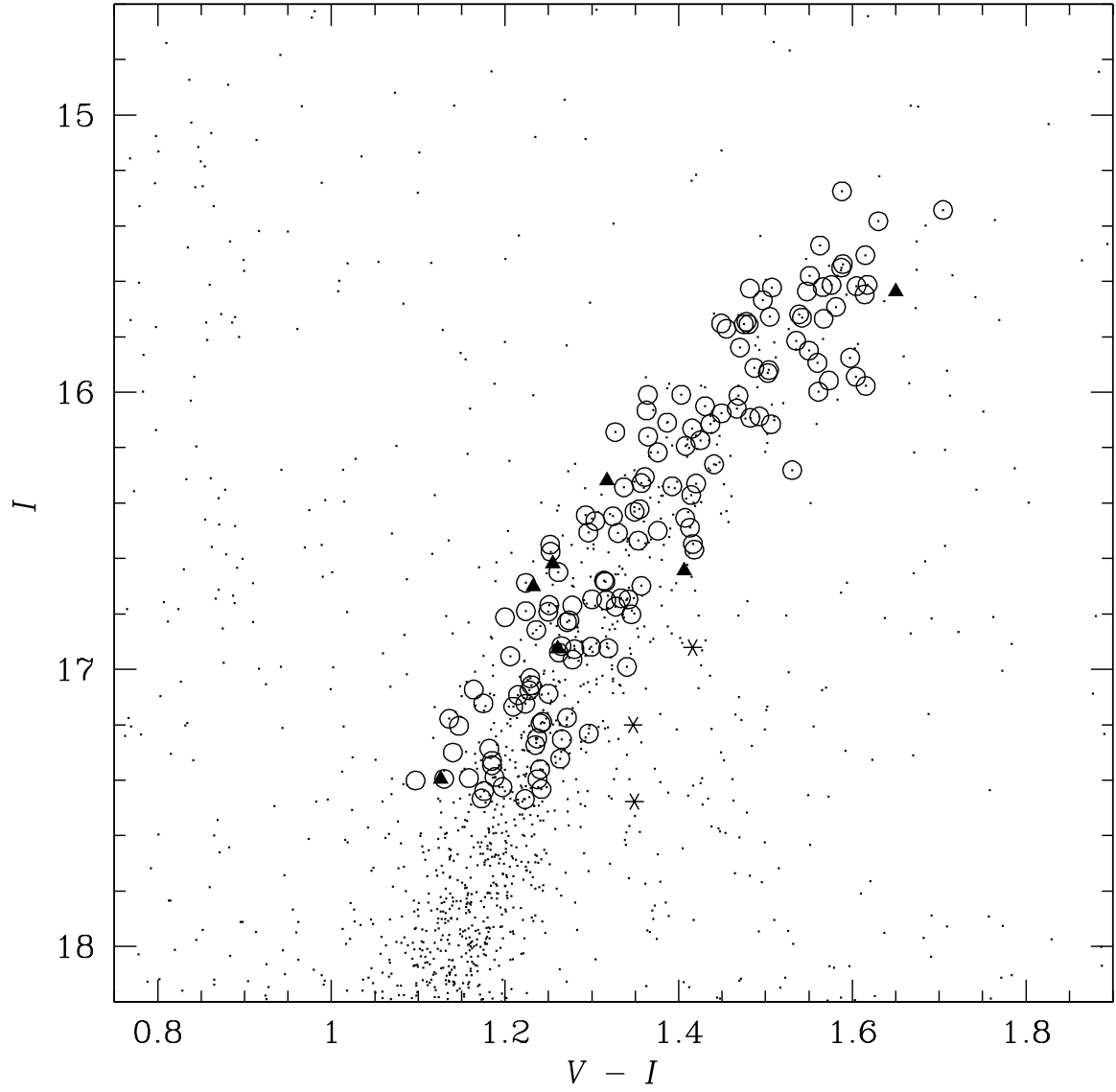}
\figcaption[CMD of Scl spectroscopic members]{Colour-magnitude data of the 148 Scl members (using the same notation as Fig.\ \ref{sclspecxy}) overlaid on the CMD for the inner $30'$ of Scl.  Stars within the tidal radius are represented as open circles, while the triangles represent extra-tidal stars.  The asterisks represent the three extra-tidal candidates rejected as Scl members based on their colours.  \label{sclspeccmd}}
\end{figure}

The sigma-clipping processes above removed six contaminating objects, two based on calcium triplet measurements and four on proper motion.  This reduced the number of possible Scl members to 151 stars, of which ten are located outside the nominal tidal radius.  Three of these extra-tidal candidates have relatively large velocity errors, which makes it difficult to be certain of membership, and colours that place them $\sim$0.05 mag to the red of the the definite Scl RGB members in the CMD (see Fig.\ \ref{sclspeccmd}).  For these reasons they were discarded from the Scl members sample.  Consequently, we have seven candidate extra-tidal members.  The characteristics of these stars are listed in Tables \ref{xtidalcandsphot} and \ref{xtidalcandsspec}.  Fig.\ \ref{sclspecxy} shows the spatial distribution of the 148 Scl members, where the closed triangles represent the seven extra-tidal stars.  Note that those stars within the tidal radius uncertainty of this limit are not considered to be extra-tidal.  To aid the reader in matching the data in Tables \ref{xtidalcandsphot} and \ref{xtidalcandsspec} with the data points in Fig.\ \ref{sclspecxy}, we have also listed their $\Delta \alpha$ and $\Delta \delta$ coordinates, and their distance from the centre of Scl, $d_{\mbox{\scriptsize Scl}}$.  The colour-magnitude distribution of the final member star list is shown in Fig.\ \ref{sclspeccmd}, where the asterisks represent the three stars rejected based on colours.  The parameters we have measured for all Sculptor members are given in Table \ref{sclmembers}.

\section{Discussion} \label{disc}
\subsection{Inner Structure}
\citet{demers80} found that the ellipticity of Scl increased with radius, which has since been confirmed in subsequent photographic plate analyses \citep{eskridge88a,eskridge88b,ih95}.  We also noted this effect.  In our analysis, the position angle of these ellipses did not significantly change with radius ($\Delta \mbox{PA} \sim 10^{\circ}$).  A possible indication of tidal distortion was the fluctuation of the central coordinates of Scl.  That is, as the ellipse radius increased, the centre shifted by approximately $10'$ North.  As such, the internal structure of Scl displays what may be the hallmarks of tidal distortion.  However, it is worth noting that a triaxial structure can also produce these chracteristics.

\citet{walcher03} described a feature extending South from the Eastern lobe of Scl (see their Fig.\ 2).  This feature was located near their tidal limit of the system ($r_t = 44'$) and they stated that it may indicate a tidal extension.  We find no evidence of this feature in the current study.  Also, \citet{walcher03} measured the tidal radius of Sculptor to be $44'$ (significantly less than previous surveys), with clear signs of tidal structure.  In contrast, we have fitted a two-component King model to Sculptor with a limiting radius of $70 - 80$ arcmin, and found that it adequately describes the system with little sign of tidal inflation.  This reinforces the \citet{tolstoy04} result, who found that Sculptor contains two separate stellar populations.

Tidal distortion may also produce kinematic substructure (such as apparent rotation) in a satellite system.  We have identified 148 Scl members through a spectroscopic analysis, and the spatial distribution of these is shown in Fig.\ \ref{sclspecxy}.  An examination of the velocities of these stars as a function of position angle revealed no significant evidence for rotation, confirming the original result by \citet{arman86}.  \citet{kleyna03} conducted a radial velocity analysis of the UMi dSph and discovered a kinematically cold population of stars towards the centre of the system, which they hypothesise may represent an underlying clump of dark matter.  To investigate the presence of kinematic substructure within Scl, we measured the offset in velocity of each star from the mean velocity of Scl, and examined this offset as a function of position.  There were no obvious kinematic substructures, however the small number of stars in the outer regions of Scl for which we have radial velocities precludes a strong result, as do the relatively large velocity errors.

\subsection{Extra-Tidal Structure}
Our analysis of the photometric survey set the upper limit for extra-tidal material to be approximately $10\%$ of the Sculptor mass.  With the spectroscopic survey, we were able to strengthen this result.  By selecting stars based on radial velocity, metallicity and proper motion, we found seven extra-tidal candidates out of a total member list of 148 objects.  These seven stars may include some residual Galactic thick disk stars, thus they correspond to an upper limit for the total mass outside the Scl tidal radius.  However, the spectroscopic survey was not complete: we obtained radial velocities for 723 of the 1408 CMD-selected stars.  Thus, by examining the number of candidates surveyed beyond the tidal radius, we find that the seven extra-tidal candidates correspond to $13.6$ stars in a complete survey of the RGB candidates.  Similarly, the density of the field stars in this CMD selection range is $\sim$$80$ stars/deg${}^2$, from which we infer that approximately 600 of the 1408 stars within the tidal radius are Sculptor members.  Therefore, assuming the extra-tidal candidates represent a stellar population with the same luminosity function as Scl, then we have measured the upper limit of extra-tidal luminous mass to be $2.3 \pm 0.6 \%$ of the Scl total, where the uncertainty represents Poisson noise.

This result can be compared to the previous study by \citet{innanen79}, who analysed 602 variable stars (mostly RR Lyrae type) in the Scl region using the catalogue of \citet{vanagt78}, described to be $75\%$ complete.  From an analysis of the stellar positions (see their Fig.\ 2) we find $\sim$$10$ of these are located beyond the nominal tidal radius of the RGB population determined in the current study, $r_t = 72.5'$.  That is, only $1-2\%$ of these stars are extra-tidal.  This result agrees with our upper mass limit for extra-tidal structure stated above.

Although there is little evidence of significant extra-tidal structure in our spectroscopic sample, it is worth refering to the work of \citet{schweitzer95}, who measured the absolute proper motion of Scl to be towards the north-east.  Four of the seven extra-tidal stars we have identified are located in this quadrant.  If these stars are members of Sculptor, then they would represent leading tidal material.  It is difficult, however, to make any definite statement in this regard based on such a small dataset.  Also, there is no correlating north-east population of extra-tidal stars in the \citet{innanen79} analysis.

\section{Conclusion}
We have conducted a combined photometric and spectroscopic survey of the Scl dSph to determine if its structure and kinematics have been influenced by the tidal field of the Galaxy.  In the first phase of the survey, we obtained photometry in two colours ($V$ and $I$) over a 10 deg${}^2$ area centred on Scl.  The dataset was complete to a magnitude of $V=20$, encompassing the Scl red giant branch.  Stars were selected in the RGB colour-magnitude range, thereby increasing the contrast of Scl stars to the foreground Galactic population.  The Sculptor structure appears to follow a two-component model, where each component is represented by the blue and red horizontal branch stars respectively.  Although there was no obvious extra-tidal structure, we determined that up to $10\%$ of the light from Scl could be outside the tidal radius and still evade photometric detection.

Hence, to further probe the Scl extra-tidal region, we obtained far red spectra for over 700 stars located in the upper-RGB region of the Scl CMD.  Stars were selected as Scl members based on three criteria: radial velocity, Ca {\sc ii} triplet strength, and proper motion.  The final Scl member list contained 148 stars, with seven located outside the tidal radius.  Thus, if Scl does possess extra-tidal structure, then it makes up less than $2.3 \pm 0.6 \%$ of the luminous mass in this system.  Therefore, we have found little sign of tidal interaction between Sculptor and the Galaxy.  This lack of tidal interaction makes it highly likely that velocity dispersion studies do provide a true measure of the dark matter mass of this dSph.

\acknowledgments
The authors would like to thank Brian Schmidt for providing the astrometric calibration program and Laura Dunn for contributing stellar population synthesis models.  MC also thanks Paul Allen for his helpful discussions on the angular correlation function, and John Norris for advice concerning the properties of the Galactic thick disk.  The authors acknowledge the 2dF observing assistance provided by Terry Bridges, and the help of Jeremy Bailey in reducing these data at the Anglo-Australian Observatory.  MC acknowledges the financial support provided by an Australian Postgraduate Award.  The authors also thank the anonymous referee for a detailed report which led to an improved paper.  This research has been supported in part by funding from the Australian Research Council under Discovery Project Grant DP034350.  This research has also made use of data obtained from the SuperCOSMOS Science Archive, prepared and hosted by the Wide Field Astronomy Unit, Institute for Astronomy, University of Edinburgh, which is funded by the UK Particle Physics and Astronomy Research Council.


\clearpage

\clearpage

\clearpage

\clearpage

\clearpage

\clearpage

\clearpage

\clearpage

\clearpage


\clearpage

\begin{sidewaystable}
\caption{Sculptor photometric observations}
\label{sculptorobservations}
\vspace{0.2cm}
\hspace{2cm}
\begin{tabular}{lcc|lcc|lcc}
\hline
\hline
 & & & & $V$ Images & & & $I$ Images & \\
Field & $\alpha$ (J2000.0) & $\delta$ (J2000.0) & Date Obs. & Exposure (s) & Seeing & Date Obs. & Exposure (s) & Seeing \\
\hline
1 & 01:05:30.3 & -32:35:05 & 2002-11-08 & $4 \times 600$ & $1.8''$ & 2002-11-03 & $4 \times 480$ & $1.7''$ \\
2 & 01:01:56.2 & -32:35:06 & 2002-11-05 & $4 \times 600$ & $2.6''$ & 2002-11-05 & $4 \times 480$ & $2.0''$ \\
3 & 00:58:21.7 & -32:35:05 & 2002-11-08 & $8 \times 300$ & $1.6''$ & 2002-11-06 & $7 \times 240$ & $1.7''$ \\
4 & 00:54:26.8 & -32:35:00 & 2002-11-08 & $4 \times 600$ & $1.5''$ & 2002-09-29 & $5 \times 500$ & $3.1''$ \\
5 & 01:05:29.8 & -33:20:06 & 2002-09-29 & $5 \times 600$ & $3.5''$ & 2002-09-29 & $5 \times 500$ & $3.2''$ \\
6 & 01:01:56.3 & -33:19:57 & 2002-09-28 & $5 \times 600$ & $2.8''$ & 2002-09-28 & $5 \times 500$ & $2.7''$ \\
7 & 00:58:21.5 & -33:19:59 & 2002-09-30 & $4 \times 600$ & $2.2''$ & 2002-10-01 & $4 \times 400$ & $1.4''$ \\
8 & 00:54:46.7 & -33:20:04 & 2002-10-01 & $4 \times 500$ & $1.9''$ & 2002-10-01 & $5 \times 400$ & $1.9''$ \\
9 & 01:05:29.6 & -34:05:09 & 2002-10-02 & $5 \times 500$ & $1.6''$ & 2002-10-02 & $5 \times 400$ & $1.7''$ \\
10 & 01:01:55.7 & -34:04:52 & 2002-10-01 & $5 \times 500$ & $1.5''$ & 2002-10-01 & $5 \times 400$ & $1.7''$ \\
11 & 00:58:22.3 & -34:04:59 & 2002-10-01 & $5 \times 500$ & $1.6''$ & 2002-10-01 & $5 \times 400$ & $1.7''$ \\
12 & 00:54:46.6 & -34:04:57 & 2002-10-02 & $4 \times 500$ & $1.9''$ & 2002-10-02 & $4 \times 400$ & $2.0''$ \\
13 & 01:05:30.4 & -34:50:07 & 2002-10-28 & $9 \times 600$ & $1.8''$ & 2002-10-28 & $5 \times 500$ & $1.8''$ \\
14 & 01:01:56.1 & -34:49:59 & 2002-10-29 & $7 \times 300$ & $1.5''$ & 2002-10-29 & $4 \times 480$ & $1.5''$ \\
15 & 00:58:21.7 & -34:49:51 & 2002-10-30 & $8 \times 300$ & $1.5''$ & 2002-10-30 & $4 \times 480$ & $1.4''$ \\
16 & 00:54:47.5 & -34:50:15 & 2002-10-31 & $9 \times 300$ & $2.6''$ & 2002-10-31 & $4 \times 480$ & $2.5''$ \\
\hline
Total & --- & --- & --- & $42100$s &  $2.0''$  & --- & $32480$s & $2.0''$ \\
\hline
\hline
\end{tabular}
\end{sidewaystable}

\begin{table}
\centering
\caption[2dF observations]{2dF observations}
\label{2dfobs}
\vspace{0.2cm}
\begin{tabular}{l|cccc}
\hline
\hline
Field & Date & Exp (sec) & Airmass & Seeing (arcsec) \\
\hline
Scl NE & 25 Sep 2003 & $5 \times 1200$ & 1.04 & 1.4 \\
Scl NW & 26 Sep 2003 & $6 \times 1200$ & 1.26 & 1.8 \\
Scl SW & 26 Sep 2003 & $6 \times 1200$ & 1.08 & 2.0 \\
Scl SE & 28 Sep 2003 & $6 \times 1200$ & 1.15 & 1.5 \\
Scl Central & 28 Sep 2003 & $6 \times 1200$ & 1.11 & 1.5 \\
NGC 1904 & 26 Sep 2003 & 750 & 1.02 & 2.0 \\
NGC 2298 & 28 Sep 2003 & 600 & 1.11 & 1.5 \\
\hline
\hline
\end{tabular}
\end{table}

\begin{table}
\centering
\caption[King Model Parameters]{King model parameters for the three Scl stellar populations.}
\label{kingpar}
\vspace{0.2cm}
\begin{tabular}{l|ccccc}
\hline
\hline
Population & Central Density & $r_c$    & $r_{\mbox{\scriptsize lim}}$    & $c$ & Background \\
           &     & (arcmin) & (arcmin) &     & (stars/arcmin${}^2$) \\
\hline
RGB & $12.0$ & $6.8 \pm 1.2$ & $72.5 \pm 4.0$ & $1.028$ & $0.0893 \pm 0.0104$ \\
RGB-a & $9.4$ & $7.6 \pm 1.4$ & $40.5 \pm 2.5$ & $0.727$ & $0.0893 \pm 0.0104$ \\
RHB & $5.9$ & $4.9 \pm 0.7$ & $35.5 \pm 2.0$ & $0.860$ & $0.0186 \pm 0.0069$ \\
BHB & $2.6$ & $7.3 \pm 1.1$ & $81.5 \pm 4.0$ & $1.048$ & $0.0030 \pm 0.0002$ \\
\hline
\hline
\end{tabular}
\end{table}

\begin{table}
\centering
\caption[BHB field-recovery statistics]{Recovery statistics for BHB stars in each field}
\label{bhbstats}
\vspace{0.2cm}
\begin{tabular}{l|ccc}
\hline
\hline
Field & Recovery ($\%$) & $\sigma_V$ (mag) & $\sigma_I$ (mag) \\
\hline
1 & $57.6 \pm 1.8$ & $0.053$ & $0.110$ \\
2 & $74.9 \pm 1.6$ & $0.077$ & $0.128$ \\
3 & $48.8 \pm 1.9$ & $0.287$ & $0.137$ \\
4 & $21.8 \pm 2.9$ & $0.152$ & $0.573$ \\
5 & $2.34 \pm 8.8$ & $0.037$ & $0.428$ \\
6 & $21.7 \pm 2.8$ & $0.058$ & $0.122$ \\
7 & $97.9 \pm 1.4$ & $0.064$ & $0.077$ \\
8 & $88.7 \pm 1.4$ & $0.048$ & $0.094$ \\
9 & $95.8 \pm 1.4$ & $0.065$ & $0.145$ \\
10 & $85.5 \pm 1.4$ & $0.103$ & $0.185$ \\
11 & $95.0 \pm 1.4$ & $0.061$ & $0.105$ \\
12 & $79.2 \pm 1.5$ & $0.058$ & $0.089$ \\
13 & $97.0 \pm 1.4$ & $0.053$ & $0.101$ \\
14 & $94.7 \pm 1.4$ & $0.059$ & $0.084$ \\
15 & $99.1 \pm 1.4$ & $0.056$ & $0.105$ \\
16 & $55.3 \pm 1.8$ & $0.074$ & $0.106$ \\
\hline
\hline
\end{tabular}
\end{table}

\begin{sidewaystable}
\caption{Candidate extra-tidal Scl stars --- photometric properties}
\label{xtidalcandsphot}
\vspace{0.2cm}
\hspace{2cm}
\begin{tabular}{l|cccccc}
\hline
\hline
Star & $\alpha$ (J2000.0) & $\delta$ (J2000.0) & $I$ & $V-I$ & $\mu_{\alpha}$ (mas/year) & $\mu_{\delta}$ (mas/year) \\
\hline
 1\_1\_240 & 01:07:22.81 & --32:21:13.99 & $16.64 \pm 0.02$ & $1.41 \pm 0.03$ & $11.0 \pm 8.4$ & $-16.4 \pm 8.3$ \\
 1\_4\_548 & 01:06:02.01 & --32:59:42.36 & $15.64 \pm 0.02$ & $1.65 \pm 0.03$ & $20.0 \pm 8.6$ & $-9.5 \pm 8.1$ \\
 1\_8\_249 & 01:04:27.87 & --32:14:43.56 & $17.40 \pm 0.02$ & $1.13 \pm 0.03$ & $11.9 \pm 8.4$ & $-13.4 \pm 8.0$ \\
 2\_3\_289 & 01:02:23.29 & --32:42:57.54 & $16.62 \pm 0.02$ & $1.26 \pm 0.03$ & $1.1 \pm 8.9$ & $-7.3 \pm 8.4$ \\
 4\_3\_158 & 00:55:37.52 & --32:41:58.09 & $16.93 \pm 0.02$ & $1.26 \pm 0.03$ &$-6.0 \pm 15.7$ & $-7.2 \pm 14.6$ \\
 13\_5\_446 & 01:03:40.85 & --35:09:10.23 & $16.32 \pm 0.02$ & $1.32 \pm 0.03$ & $2.9 \pm 8.8$ & $-13.7 \pm 8.2$ \\
 16\_2\_265 & 00:55:21.21 & --34:42:12.65 & $16.70 \pm 0.02$ & $1.23 \pm 0.03$ & $28.8 \pm 14.0$ & $-24.2 \pm 13.2$ \\
\hline
\hline
\end{tabular}
\end{sidewaystable}

\begin{table}
\begin{center}
\caption{Candidate extra-tidal Scl stars --- spectroscopic properties}
\label{xtidalcandsspec}
\vspace{0.2cm}
\begin{tabular}{l|cccccc}
\hline
\hline
Star & $v_r$ & [Fe/H] & $\Delta \alpha$ & $\Delta \delta$ & $d_{\mbox{\scriptsize Scl}}$ \\
 & (km s${}^{-1}$) &  & (${}^{\circ}$) & (${}^{\circ}$) & (${}^{\circ}$) & \\
\hline
 1\_1\_240 & $89.1 \pm 1.9$ & $-1.67 \pm 0.17$ & $1.50$ & $1.35$ & $2.02$ \\
 1\_4\_548 & $84.0 \pm 0.7$ & $-1.89 \pm 0.11$ & $1.22$ & $0.71$ & $1.42$ \\
 1\_8\_249 & $86.6 \pm 7.8$ & $-1.66 \pm 0.08$ & $0.90$ & $1.46$ & $1.72$ \\
 2\_3\_289 & $91.3 \pm 7.1$ & $-1.70 \pm 0.21$ & $0.47$ & $0.99$ & $1.10$ \\
 4\_3\_158 & $95.8 \pm 11.8$ &  $-1.86 \pm 0.36$ & $-0.94$ & $1.01$ & $1.38$ \\
 13\_5\_446 & $126.6 \pm 3.8$ & $-1.66 \pm 0.29$ & $0.73$ & $-1.45$ & $1.62$ \\
 16\_2\_265 & $112.9 \pm 3.7$ & $-1.40 \pm 0.22$ & $-1.00$ & $-1.00$ & $1.41$ \\
\hline
\hline
\end{tabular}
\end{center}
\end{table}

\begin{sidewaystable}
\caption{Sculptor members}
\label{sclmembers}
\vspace{0.2cm}
\hspace{2cm}
\begin{tabular}{l|cccccc}
\hline
\hline
Star & $\alpha$ (J2000.0) & $\delta$ (J2000.0) & $V$ & $V-I$ & $v_r$ (km/s) & [Fe/H] \\
\hline
 16\_2\_265 &   00:55:21.20 & --34:42:12.6 & $17.934 \pm 0.005$ & $1.232 \pm 0.009$ & $112.9 \pm 3.7$ &  $-1.40 \pm 0.22$ \\
 12\_2\_332 &   00:55:21.98 & --34:01:42.4 & $18.498 \pm 0.006$ & $1.097 \pm 0.008$ & $118.7 \pm 14.8$ & $-2.38 \pm 0.17$ \\
 8\_3\_296 &    00:55:23.44 & --33:24:09.9 & $18.639 \pm 0.006$ & $1.173 \pm 0.008$ & $101.7 \pm 6.7$ &  $-1.78 \pm 0.26$ \\
 4\_3\_158 &    00:55:37.52 & --32:41:58.0 & $18.186 \pm 0.006$ & $1.260 \pm 0.007$ & $95.8 \pm  11.8$ & $-1.86 \pm 0.36$ \\
 8\_4\_235 &    00:55:55.74 & --33:33:53.3 & $18.351 \pm 0.005$ & $1.146 \pm 0.007$ & $95.1 \pm  19.0$ & $-2.00 \pm 0.19$ \\
 12\_3\_139 &   00:56:12.09 & --34:16:17.7 & $18.012 \pm 0.004$ & $1.199 \pm 0.006$ & $114.0 \pm 2.5$ &  $-2.27 \pm 0.21$ \\
 7\_7\_306 &    00:56:20.23 & --33:09:20.9 & $18.551 \pm 0.009$ & $1.158 \pm 0.012$ & $108.6 \pm 19.4$ & $-1.24 \pm 0.13$ \\
 11\_8\_435 &   00:56:46.94 & --33:48:20.6 & $17.235 \pm 0.003$ & $1.480 \pm 0.005$ & $116.2 \pm 3.4$ &  $-2.06 \pm 0.10$ \\
 3\_5\_175 &    00:57:15.38 & --32:59:28.4 & $18.046 \pm 0.006$ & $1.277 \pm 0.010$ & $89.9 \pm  5.3$ &  $-1.46 \pm 0.09$ \\
 7\_5\_285 &    00:57:19.67 & --33:36:38.0 & $18.525 \pm 0.008$ & $1.129 \pm 0.010$ & $126.6 \pm 2.5$ &  $-1.83 \pm 0.20$ \\
\hline
\hline
\end{tabular}

{\small {\sc Note:} Table \ref{sclmembers} is presented in its entirety in the electronic edition of the Astronomical Journal.  A portion is shown here for guidance regarding its form and content.  The uncertainties for $V$ and $(V-I)$ are relative only, and do not include those from the photometric calibration.}
\end{sidewaystable}

\end{document}